\documentclass[12pt,reqno]{amsart}
\usepackage{amsmath,amssymb,amsfonts,amsthm}
\usepackage{graphicx}
\usepackage{geometry}
\usepackage{color}
\usepackage{enumitem}
\usepackage{mathtools}
\usepackage{setspace}
\usepackage{bbm}
\usepackage{natbib}
\newtheorem{theorem}{Theorem}
\theoremstyle{plain}
\newtheorem{assumption}{Assumption}

\newtheorem{lemma}{Lemma}
\newtheorem{proposition}{Proposition}
\theoremstyle{definition}
\newtheorem{definition}{Definition}
\newtheorem{example}{Example}
\AddToHook{env/example/begin}{\pushQED{\qed}}
\AddToHook{env/example/end}{\popQED}
\theoremstyle{remark}

\newcommand{\real}{\mathbb{R}}
\newcommand{\supp}{\mathrm{supp}}

\newcommand{\Gr}{\text{\normalfont Gr}}
\newcommand{\co}{\text{\normalfont co}}				\newcommand{\clco}{\overline{\text{\normalfont co}}}		\newcommand{\cl}{\text{\normalfont cl}}							
\newcommand{\weakconv}{\rightsquigarrow}

\newcommand\independent{\protect\mathpalette{\protect\independenT}{\perp}}
\def\independenT#1#2{\mathrel{\rlap{$#1#2$}\mkern2mu{#1#2}}}
\DeclareMathOperator*{\argmax}{arg\,max}
\DeclareMathOperator*{\argmin}{arg\,min}
\numberwithin{equation}{section}
\newenvironment{continued}[1][continued]{\pushQED{\qed}\begin{trivlist}
\item[\hskip \labelsep {\bfseries #1}]}{\popQED\end{trivlist}}

\begin{document}

\title[Indirect inference in incomplete models]{Indirect inference in incomplete and partially identified econometric models}
\author{Lixiong Li \and Marc Henry}
\address{Johns Hopkins and Penn State}

\thanks{The first version is of July 14, 2026. This version is of October 21, 2026. The authors gratefully acknowledge excellent research assistance from Onil Boussim. Corresponding author: Lixiong Li: \texttt{lixiong.li@jhu.edu}, Department of Economics, Johns Hopkins University, Wyman Park Building 5th Floor, 3100 Wyman Park Drive, Baltimore, MD 21211.}

\begin{abstract}
We extend the indirect inference principle to structural econometric models that are incomplete, in the sense that a single value of the structural parameter may be compatible with multiple distributions of the observed endogenous variables, so that the structural parameter is typically only partially identified. We define the binding function as the smallest norm of the auxiliary estimating equations over the set of data generating processes compatible with the structural model. Hence, we obtain a confidence region by inverting a test of the hypothesis that the binding function vanishes at the pseudo-true value. A dual characterization based on random set theory makes the binding function computationally tractable. We define the auxiliary model conditionally on the covariates to remove its dependence on the unknown covariate distribution and we show that the binding function is Lipschitz and Hadamard directionally differentiable, so that the numerical delta method delivers valid critical values. We show that the method provides significant computational advantages without relinquishing informativeness in a simulation experiment based on an entry game with~$100$ potential entrants.

\vskip20pt


\vskip10pt


\end{abstract}

\maketitle

\section*{Introduction}

Many structural econometric models are incomplete: for a given value of the structural parameter, and given realizations of the exogenous and latent variables, the model may fail to deliver a unique prediction for the endogenous outcomes. Leading examples include game theoretic models with multiple equilibria, as in empirical entry models \citep{BR:91,Tamer:2003,CT:2009,CMT:2021}, English auction models under weak assumptions on bidding behavior \citep{HT:2003,CR:2017}, models of network formation \citep{dPRST:2018,Chandrasekhar:2016}, and discrete choice models with heterogeneous and unobserved choice sets \citep{BCMT:2021}. Rather than imposing an ad hoc equilibrium selection mechanism or other completing assumptions, one may take the incompleteness seriously, at the cost of losing point identification of the structural parameter. The econometrics of such partially identified models is by now well developed. See \cite{HR:2017}, \cite{CS:2018} and \cite{Molinari:2020} for recent surveys.

The sharp identified region of an incomplete structural model is characterized by the requirement that the true distribution of outcomes be compatible with the distribution of the latent variables and the support restrictions imposed by the model. This compatibility can be formulated as an optimal transport problem \citep{GH:2006,EGH:2010} or, equivalently, as a collection of inequalities, due to Artstein, indexed by subsets of the outcome space \citep{GH:2011,BMM:2011,CR:2017e}. Inference can then in principle proceed with moment inequality testing. In practice, however, the number of Artstein inequalities grows exponentially with the cardinality of the outcome space, and although core determining classes \citep{GH:2011,LW:2017,luo2024selecting} can dramatically reduce their number, the remaining collection is often still too large for existing many-inequality methods. Practitioners therefore commonly select a manageable subset of inequalities, and thereby incur a loss of sharpness that is difficult to evaluate or even interpret.

This paper proposes instead to extend the scope of indirect inference to incomplete structural models. Since \cite{Smith:93}, \cite{GMR:93} and \cite{GT:96}, indirect inference has been a common inference method for complete structural models that are easy to simulate but whose likelihood is intractable. The method replaces the structural likelihood with a descriptive auxiliary model, possibly misspecified, whose pseudo-true parameter is easy to estimate. The structural parameter is then recovered by matching the auxiliary estimates computed from the observed data with those computed from data simulated from the structural model, through the so-called binding function. Indirect inference has been applied, among many other examples, to the estimation of labor income risk \citep{Guvenen:2014}, dynamic industry equilibrium \citep{Collard-Wexler:2013}, equilibrium search models of wage dynamics and dispersion \citep{Fontaine:2014,Bagger:2019}, models of media bias \citep{Yurukoglu:2017}, and structural general equilibrium models of trade and economic geography estimated with a mix of targeted moments and auxiliary regressions \citep{Tintelnot:2017,Gaubert:2018}.

Our procedure starts with an auxiliary model described by a vector of estimating equations. These equations depend on a finite-dimensional auxiliary parameter and have zero expectation at a unique pseudo-true value under the true data generating process. In a complete structural model, the usual binding function maps each structural parameter to the auxiliary parameter implied by the model. At any structural parameter in the identified set, the implied auxiliary parameter equals the pseudo-true value. A confidence region can therefore be constructed by inverting a test of this equality.  In an incomplete model, however, each structural parameter is compatible with a set of data generating processes. 
We therefore introduce a new binding function that extends the matching principle to incomplete structural models. Specifically, we define the binding function as the smallest norm of the expected auxiliary estimating equations over all compatible data generating processes. At any structural parameter in the identified set, this set includes the true data generating process. Because the auxiliary estimating equations have zero expectation at the pseudo-true value under that process, the binding function must also vanish there. We obtain a confidence region by inverting a test of this zero restriction.

In addition to introducing this binding function for incomplete structural models, we resolve three complications that arise in its computation and use for inference.
\begin{enumerate}
\item First, the binding function, hence also the test statistic, is defined as the value of an infinite dimensional optimization program over sets of conditional distributions. We derive a dual characterization, based on the theory of random sets, that reduces its computation to a finite dimensional concave maximization problem, and allows the computation of analytical and numerical derivatives.
\item Second, in the absence of covariates, the binding function depends only on known quantities, such as the distribution of the latent variables. With covariates, the binding function also depends on the unknown distribution of the exogenous covariates. We therefore define the auxiliary model conditionally on the covariates, and define the binding function as a function of the known empirical distribution of the covariates.
\item Third, as is customary in partially identified models, the binding function is not Hadamard differentiable in the auxiliary parameter at all structural parameter values. We show that it is Lipschitz and Hadamard directionally differentiable, and we apply the numerical delta method of \cite{HL:2018} to compute valid critical values.
\end{enumerate}

We illustrate the computational feasibility and informativeness of our approach with a simulation experiment based on an entry game with \(100\) potential entrants whose profits depend on firm-specific covariates and idiosyncratic payoff shocks.
Here,  a sharp characterization of the identified sets based on Artstein inequalities or core determining classes is computationally prohibitive. Our indirect inference procedure exploits the low dimension of the auxiliary estimating equations and the structure of the equilibrium correspondence. With a sample size of $5{,}000$ and $500$ resampling draws for computing critical values, testing one candidate parameter takes approximately $1$ to $3$ seconds on a standard desktop computer with a single CPU thread. The auxiliary model also makes the loss of identifying information transparent and interpretable: identification uses only the features of the data distribution captured by the auxiliary estimating equations. Auxiliary-model choice therefore governs this information loss, and a specification that captures relevant data variation can yield more informative confidence regions. In our experiment, augmenting the auxiliary regressions with rival firms’ characteristics substantially shortens the projected confidence intervals for the competition parameters.

\subsection*{Related literature}

The paper contributes to the literature on indirect inference and to the literature on inference in partially identified models.

The indirect inference method, in the sense of inference based on intermediate statistics, is surveyed by \cite{Turnbull:2004}. Within this tradition, \cite{Li:2010} proposes ordinary least squares auxiliary models for structural microeconometric models; \cite{Calvet:2015} construct auxiliary models from simpler equilibrium concepts in simpler economies, with an application to the long run risk model of \cite{Bansal:2004}; \cite{Blasques:2018} incorporate prior information on the structural parameter through a penalized Wald-type indirect inference criterion; \cite{Antoine:2022} develop identification-robust indirect inference for a subvector of structural parameters when the remaining parameters are calibrated, with an application to DSGE models; and \cite{Chaudhuri:2018} treat indirect inference with endogenously missing exogenous variables, a setting whose incomplete-model version, with weaker assumptions on the sample selection mechanism, fits the framework of the present paper. \cite{Sauer:2021} note that indirect inference has traditionally been avoided in models with discrete choices because of the non smoothness of the objective function, and propose an importance sampling solution; \cite{FOZ:2019} develop the asymptotic theory of indirect inference with non smooth criterion functions. Since the auxiliary model is generically misspecified, the analysis of pseudo-true values and quasi-maximum likelihood \citep{White:94}, as well as finite sample concentration results for misspecified likelihoods \citep{Spokoiny:2015}, are directly relevant. \cite{Deaton:95} discuss the estimation of nonlinear rational expectations commodity price models with unobservable state variables, an early candidate application of simulation based indirect methods.

On the partial identification side, beyond the characterizations of sharp identified regions discussed above, the duality we derive for the binding function is related to the dual approach to inference of \cite{Kaido:2016}, and our use of support functions and selection expectations builds on \cite{BMM:2011,BMM:2012} and \cite{Molchanov:2017}. 
We use these tools to develop indirect inference through a potentially misspecified auxiliary model, accounting for estimation of its pseudo-true parameter in the nonsmooth binding function. 
Our formulation defines both the auxiliary target and the binding function conditional on the observed covariate design.
The lack of full differentiability of the binding function also connects the paper to the literature on inference on directionally differentiable functionals \citep{Shapiro:90,Shapiro:91,BS:2000,FS:2019,HL:2018}. Robust Bayesian approaches to set identified models are developed in \cite{GK:2021} and \cite{GKR:2021}. Estimation based on matching distributional features also appears in \cite{AB:2021} and, with the Wasserstein distance as the discrepancy, in \cite{Robert:2019}. 

On the analysis of incomplete models, this work is closely related to \cite{KM:2022}. Their strategy is to replace the true data generating process $P_0$ (or its empirical counterpart $\hat P_n$) with the pseudo-maximum likelihood projection $P_{\hat\eta}$ of the data onto the model. This allows them to analyze cases where the incomplete structural model is misspecified, which we do not pursue here.

\subsection*{Notation and preliminaries}

Throughout, $p$ denotes the density associated with a distribution $P$ when there is no ambiguity. For probability measures $P$ on $\mathcal Y$ and $Q$ on $\mathcal U$, $\mathcal M(P,Q)$ denotes the set of probability measures on $\mathcal Y\times\mathcal U$ with marginals $P$ and $Q$. For a measurable correspondence $\Gamma$ from $\mathcal U$ to $\mathcal Y$, $\mbox{Sel}^0(\Gamma)$ denotes the set of its measurable selections, and $\mbox{Sel}^1(\Gamma,Q)$ the subset of those that are $Q$-integrable. The support function of a set $A\subseteq\real^d$ in direction $a\in\real^d$ is denoted $h(a,A):=\inf_{y\in A}a^\top y$ (note the choice of the lower support function, i.e., infimum rather than supremum, throughout). $\mathcal S$ denotes the closed unit ball in $\real^{d_s}$, $\|\cdot\|$ the Euclidean norm of a vector and the operator norm it induces on matrices, $I_d$ the identity matrix of order $d$, and, for a metric space $\mathbb D$, $\mbox{BL}_1(\mathbb D)$ the class of real valued functions on $\mathbb D$ bounded by~$1$ and with Lipschitz constant bounded by~$1$; the argument is $\real^{d_\beta}$ when the bounded Lipschitz distance is applied to random vectors in $\real^{d_\beta}$, and $\real$ when it is applied to real valued statistics. For an integrably bounded, nonempty compact valued random set $\bf\Gamma$ in $\real^d$, induced by a measurable correspondence $\Gamma$ and a distribution $Q$, we write
\begin{eqnarray*}
{\bf E}{\bf\Gamma} & := & \clco\left\{ \mathbb E_Q\left[\gamma(U)\right]:\;\gamma\in\mbox{Sel}^1(\Gamma,Q) \right\}
\end{eqnarray*}
for its convexified selection expectation, i.e., the closed convex hull of its Aumann expectation \citep[Section 2.1]{Molchanov:2017}. When $Q$ is non atomic, the Aumann expectation is convex by the convexity theorem of \cite{Richter:63} and \cite{Aumann:65}. Under the nonempty compact valued and integrably bounded conditions just stated, it is also compact, and ${\bf E}{\bf\Gamma}$ reduces to the selection expectation in the usual sense. Weak convergence is denoted $\weakconv$.

\subsection*{Overview}

Section~\ref{sec:model} defines the class of incomplete structural models under consideration, presents leading examples, and defines the sharp identified region. Section~\ref{sec:ii} introduces the auxiliary model and the binding function, establishes the dual characterization of the latter, and shows that it is Lipschitz and Hadamard directionally differentiable in the auxiliary parameter. Section~\ref{sec:inference} presents the confidence region for the structural parameter, based on the numerical delta method, and establishes its asymptotic validity. Section~\ref{sec:simulation} reports Monte Carlo evidence. Section~\ref{sec:discussion} discusses extensions and the relation to alternative proposals. Proofs are collected in the appendix, together with a generic construction of the resampling scheme and the extension of the results to weighted auxiliary estimating equations with estimated weights.

\section{Incomplete model}
\label{sec:model}

\subsection{Theoretical structural model}

We restrict our attention to the class of incomplete parametric structural models introduced in \cite{Jovanovic:89}. The vector of variables of interest $(Y,X,U)\in\mathcal Y \times \mathcal X \times \mathcal U$ satisfies the support constraint $Y \in \Gamma(U\vert X;\theta_1)\subseteq \mathcal Y$, and the probability distribution of $U$ conditional on possible realizations $x$ of $X$ is given by $Q_{U\vert x;\theta_2}$. The object of inference is the finite dimensional parameter $\theta:=(\theta_1,\theta_2)\in\Theta$. Both vectors of variables $Y$ and $X$ are observed, in the sense that the available data consist of an i.i.d. sample $((Y_1,X_1),\ldots,(Y_n,X_n))$. Variables in vector $U$ are unobserved. Variables in vector $X$ are exogenous, in the sense that the conditional distribution $Q_{U\vert x;\theta_2}$ is fixed a priori. This includes $U\independent X$ as a special case. All endogenous (i.e., non exogenous) variables are subsumed in the vector $Y$.

The incompleteness of the model is reflected in the fact that multiple values of the endogenous variables may be consistent with a single value of the exogenous and unobserved variables: the set $\Gamma(u\vert x;\theta_1)\subseteq \mathcal Y$ may not be a singleton for all $(u,x)\in\mathcal U\times\mathcal X$, so that the model fails to produce a unique prediction.

\subsection{Examples}

\begin{example}[Market entry with multiple equilibria]
\label{ex:entry}
In the two firm entry game of \cite{BR:91} and \cite{Tamer:2003}, the outcome $Y\in\{0,1\}^2$ records the entry decisions of both firms in a market with characteristics $X$, and $U$ collects the latent profit shifters of the two firms. For a given value of the payoff parameters $\theta_1$, the correspondence $\Gamma(U\vert X;\theta_1)$ collects the pure strategy Nash equilibria of the game; in the region of realizations of $(U,X)$ where the game admits multiple equilibria, $\Gamma$ is multi-valued, and the model is silent on which equilibrium is selected. \cite{CT:2009} and \cite{CMT:2021} develop empirical models of airline markets in this vein.
\end{example}

\begin{example}[Incomplete model of English auctions]
\label{ex:auctions}
In the incomplete model of English auctions of \cite{HT:2003}, further studied by \cite{CR:2017}, bidders' valuations $U$ are drawn from a distribution known up to $\theta_2$, and the vector of final observed bids $Y$ is only assumed to satisfy weak behavioral restrictions: no bidder bids above their valuation, and no bidder allows an opponent to win at a price they would have been willing to beat. The set of bid vectors consistent with these restrictions, given valuations and auction covariates, defines the correspondence $\Gamma(U\vert X;\theta_1)$.
\end{example}

\begin{example}[Network formation]
\label{ex:networks}
In strategic models of network formation, the observed network $Y$ is assumed to be pairwise stable given agents' characteristics $X$ and latent preference shocks $U$. Pairwise stable networks typically exist but are not unique, so that the set of pairwise stable networks defines a multi-valued correspondence $\Gamma(U\vert X;\theta_1)$; see \cite{dPRST:2018} and the survey by \cite{Chandrasekhar:2016}.
\end{example}

\subsection{Sharp identified region}
\label{sec:sharp}

The sample $((y_1,x_1),\ldots,(y_n,x_n))$ of observed data is assumed to be a collection of realizations from a random vector $(Y,X)$ with true distribution $P_0:=P_{Y\vert X,0}\times P_{X,0}$. The model stipulates that $P_{Y\vert X,0}$ is an element of a set $\mathcal P_\theta$ of regular conditional distributions\footnote{With a slight abuse of terminology and notation, we refer to a version of a conditional distribution as {\em the} conditional distribution.} defined as follows.

\begin{definition}[Structural model]
\label{def:DGP}
For each $\theta=(\theta_1,\theta_2)\in\Theta$, $\mathcal P_\theta$ is the set of Markov kernels $P_{Y\vert x}$ with source $\supp(X)$ and target $\mathcal Y$, such that there exists a Markov kernel $\pi_x$ with source $\supp(X)$ and target $\mathcal Y\times\mathcal U$, that satisfies the following constraints for all $x\in\supp(X)$:
\begin{enumerate}
\item Support restriction: $\pi_x \left( Y\in\Gamma(U \vert x;\theta_1)\subseteq\mathcal Y \right)=1$.
\item Marginal restrictions: $\pi_x$ has marginals $P_{Y\vert x}$ and $Q_{U\vert x;\theta_2}$.
\end{enumerate}
\end{definition}

Compatibility between the structural model of Definition~\ref{def:DGP} and the true data generating process is defined as the fact that $P_{Y\vert X,0}$ is an element of $\mathcal P_\theta$. Due to the incompleteness of the model, for any given $\theta$, the structural model can generate multiple predictions for the process generating the observed data, i.e., $\mathcal P_\theta$ may not be a singleton. Conversely, any given true data generating process $P_{Y\vert X,0}$ may be compatible with the structural model, i.e., $P_{Y\vert X,0}\in\mathcal P_\theta$, for more than one value of the parameter $\theta\in\Theta$. Hence the parameter vector $\theta$ may be only partially identified. The sharp identified region $\Theta_I$ is defined as the set of values of the parameter $\theta$ such that the model is compatible with the true data generating process.

\begin{definition}[Sharp identified region]
\label{def:idset}
The sharp identified region is defined as
\begin{eqnarray*}
\Theta_I := \left\{\theta\in\Theta:\;\exists P_{Y\vert X}\in\mathcal P_\theta \mbox{ such that }
P_{Y\vert x}=P_{Y\vert X=x,0}\;\mbox{ for } P_{X,0}\mbox{-almost every } x\right\}.
\end{eqnarray*}
\end{definition}

We assume that the structural model specification is non trivial in the sense that there exists $\theta\in\Theta$ such that $\mathcal P_\theta \ne \emptyset$. In other words, the sharp identified region is non empty for at least one true data generating process. However, the sharp identified region may be empty for some true data generating process $P_0$, in which case the structural model is incompatible with the data, and should be rejected.

\section{Indirect inference}
\label{sec:ii}

\subsection{Outline of the inference methodology}

The essence of indirect inference is to use an auxiliary econometric model for the observed data generating process that may not be correctly specified, but provides a good approximation of the latter. To see how this auxiliary model is used to conduct inference on the structural parameter, start with the special case without covariates, and where the set of data generating processes predicted by the structural model is a singleton, i.e., $\mathcal P_\theta=\{P_\theta\}$. The identified set of Definition~\ref{def:idset} is then $\Theta_I=\{\theta\in\Theta:\;P_\theta=P_{Y,0}\}$.

The auxiliary model can be a (potentially misspecified) statistical model $\{P_\beta:\;\beta\in\mathbb B_0\subseteq\mathbb R^{d_\beta}\}$. Suppose that there is a unique pseudo-true value $\beta_0$ which maximizes the pseudo-likelihood $\mathbb E_{P_0}\ln p_\beta(Y)$, where $p_\beta$ is the density of $P_\beta$. Following the terminology of \cite{GMR:93}, call $\theta\mapsto \argmax_\beta\mathbb E_{P_\theta}\ln p_\beta(Y)$ the binding function, assuming, for the sake of this heuristic discussion, that the maximizer is unique. By construction, at any $\theta$ in the identified set $\Theta_I$, the binding function takes the value $\beta_0$. A valid confidence region for the structural parameter $\theta$ can therefore be obtained by inverting a test of the hypothesis that the binding function takes the value $\beta_0$.

More generally, the auxiliary model may be defined by $\mathbb E_{P_0}s(Y;\beta_0)=0$ for a unique pseudo-true value $\beta_0$, where the vector function $s(y;\beta)$ may or may not be a score function. With some abuse of terminology, we also call binding function the score version
$L(\theta,\beta):=\| \mathbb E_{P_\theta}s(Y;\beta) \|,$
so that $L(\theta,\beta_0)=0$ when $\theta\in\Theta_I$ by construction. Again, a valid confidence region can be obtained by inverting a test of $L(\theta,\beta_0)=0$.

Extend the reasoning to incomplete models, where the set $\mathcal P_\theta$ of data generating processes predicted by the model is no longer a singleton. If $\theta\in\Theta_I$, i.e., $P_{Y,0}\in\mathcal P_\theta$, then there is at least one data generating process in the predicted set $P\in\mathcal P_\theta$ satisfying $\mathbb E_Ps(Y;\beta_0)=0$. Hence, defining the binding function as
\begin{eqnarray}
\label{eq:bind}
L(\theta,\beta) & := & \inf \left\{ \| \mathbb E_P \; s(Y;\beta) \|: \; P\in\mathcal P_\theta\right\},
\end{eqnarray}
a valid confidence region can again be obtained by inverting a test of $L(\theta,\beta_0)=0$.

We perform the test of $L(\theta,\beta_0)=0$ with the test statistic $L(\theta,\hat\beta)$, where the asymptotic distribution of the estimator $\hat\beta$ can be easily approximated, based on the analytical expression of the limiting distribution or by simulation. The point of indirect inference is precisely to allow easy approximation of the distribution of $\hat\beta$ by a suitable choice of the auxiliary model. Critical values are then obtained using the delta method. However, three major complications arise.
\begin{enumerate}
\item First, the binding function, hence also the test statistic, are defined as solutions to an infinite-dimensional optimization program. We propose a dual characterization that allows the computation of analytical and numerical derivatives.
\item Second, in the case without covariates, the binding function depends only on known quantities, such as the distribution $Q_{U\vert X}$ of latent variables. In the case with covariates, however, the binding function also depends on the distribution $P_{X,0}$ of exogenous covariates. We therefore propose to use a conditional auxiliary model and to define the binding function accordingly, as a function of the known empirical distribution $P_{X,n}$ of the covariates.
\item Third, as is customary in partially identified models, the binding function is not Hadamard differentiable with respect to $\beta$ for all $\theta$. We show that the binding function is Lipschitz and Hadamard directionally differentiable, and we apply the numerical delta method of \cite{HL:2018} to compute critical values.
\end{enumerate}

\subsection{Auxiliary model, estimator and resampling scheme}
\label{sec:aux}

Write~$P_0=P_{Y\vert X,0}\times P_{X,0}$ for the true data generating process and~$P_{X,n}$ for the empirical distribution associated with the sample~$(X_1,\ldots,X_n)$ of covariates. We define the auxiliary model conditionally on the covariates to bypass issues associated with nonparametric estimation of~$P_{X,0}$ within the indirect inference procedure below. The auxiliary specification has three ingredients, which we introduce in turn: a vector of estimating equations defining the pseudo-true value; an estimator sequence for the pseudo-true value; and a resampling or simulation scheme approximating the distribution of the estimator. 

\begin{assumption}[Auxiliary model]
\label{ass:aux}
There is a vector function~$(y,x,\beta)\mapsto s(y\vert x;\beta)$ from~$\mathcal Y\times\mathcal X\times \mathbb B_0$ with~$\mathbb B_0\subseteq\mathbb R^{d_\beta}$ to~$\mathbb R^{d_s}$ such that for each $n\geq1$, there is, almost surely with respect to the distribution of~$(X_i)_{i=1}^\infty$, a unique~$\beta_n$ such that
\begin{eqnarray}
\label{eq:aux}
\mathbb E_{P_{Y\vert X,0}\times P_{X,n}}\left[ s(Y\vert X;\beta_n)\right] \; = \; \frac{1}{n}\sum_{i=1}^n\mathbb E\left[ s(Y\vert X_i;\beta_n)\; \vert\; X_i\right] & = & 0.
\end{eqnarray}
\end{assumption}

Assumption~\ref{ass:aux} defines the pseudo-true value~$\beta_n$ relative to the finite covariate design. The function~$s$ may, but need not, be the score of a parametric likelihood, and the auxiliary model may be misspecified. Only existence and uniqueness of the solution~$\beta_n$ is required. In the examples below, uniqueness follows from one of two structures: either~$\beta\mapsto n^{-1}\sum_i\mathbb E[s(Y\vert X_i;\beta)\vert X_i]$ is affine in~$\beta$ with nonsingular Jacobian, as for least squares and just-identified instrumental variable scores, or it is the gradient of a strictly concave criterion, as for quasi-likelihood and pseudo-likelihood scores.

The auxiliary model is chosen in such a way that an estimator sequence~$\hat\beta_n$ is readily available and its distribution can be approximated. We first assume that the estimator sequence~$\hat\beta_n$ is well behaved asymptotically.

\begin{assumption}[Auxiliary estimator]
\label{ass:est}
There exist sequences~$r_n\rightarrow\infty$, ~$\hat\beta_n$ with values in~$\mathbb B_0$, and a Borel measurable, separable and tight random vector~$\mathbb Z_0$ with values in~$\real^{d_\beta}$, such that
\begin{eqnarray*}
\sup_{f\in\mbox{\tiny BL}_1(\real^{d_\beta})}\left| \mathbb E\left[ f( r_n(\hat\beta_n-\beta_n)) \vert (X_i)_{i=1}^n \right]
-\mathbb E\left[ f(\mathbb Z_0) \vert (X_i)_{i=1}^n \right] \right| \rightarrow0
\end{eqnarray*}
almost surely with respect to the distribution of~$(X_i)_{i=1}^\infty$.
\end{assumption}

In regular cases, $r_n=\sqrt n$ and~$\mathbb Z_0$ is a centered Gaussian vector conditionally on the covariates, but the bounded Lipschitz formulation of Assumption~\ref{ass:est}, conditional on the covariate sequence, accommodates covariate-dependent and non-Gaussian limits. 

Second, we assume that the random element~$\mathbb Z_0$ can be suitably approximated by~$\mathbb Z_n^\ast$ obtained by simulation or resampling.

\begin{assumption}[Resampling]
\label{ass:boot}
The sequence~$\mathbb Z_n^\ast$ of random vectors in~$\real^{d_\beta}$ satisfies
\begin{eqnarray*}
\sup_{f\in\mbox{\tiny BL}_1(\real^{d_\beta})}\left| \mathbb E\left[ f( \mathbb Z_n^\ast) \vert (X_i,Y_i)_{i=1}^n \right]
-\mathbb E\left[ f(\mathbb Z_0) \vert (X_i)_{i=1}^n \right] \right| \rightarrow0
\end{eqnarray*}
in probability.
\end{assumption}

The left-hand side in Assumption~\ref{ass:boot} is a measurable function of the data $(X_i,Y_i)_{i=1}^n$, and convergence in probability is with respect to their joint distribution. This formulation accommodates resampling distributions that depend on estimated nuisance quantities, such as the covariance estimator of Appendix~\ref{sec:cross-fit}. The sequence~$\mathbb Z_n^\ast$ can be generically constructed using residualized estimating equations and cross-fitting as we explain in Appendix~\ref{sec:cross-fit}, where Lemma~\ref{lem:crossfit} gives primitive conditions for Assumption~\ref{ass:boot}.

\subsection{Binding function}

We define a feasible analogue of the binding function in~(\ref{eq:bind}), where the unknown covariate distribution $P_{X,0}$ is replaced with the empirical distribution $P_{X,n}$.

\begin{definition}[Binding function]
\label{def:bind-n}
Define the binding function for any pair $(\theta,\beta)\in\Theta\times \mathbb B_0$:
\begin{eqnarray*}
L_n(\theta,\beta) & := & \inf \left\{ \left\|
\frac1n\sum_{i=1}^n\mathbb E_{P_{Y\vert X}} \left[ s(Y\vert X_i;\beta)\; \vert\; X_i\right] \right\|
: \; P_{Y\vert X} \in \mathcal P_\theta \right\}.
\end{eqnarray*}
\end{definition}

First we show that the binding function $L_n(\theta,\beta)$ is indeed equal to zero at $\beta=\beta_n$ for any $\theta\in\Theta_I$. This guarantees that we obtain a valid confidence region for each element of $\Theta_I$ by inverting a test of the hypothesis $L_n(\theta,\beta_n)=0$.

\begin{proposition}
\label{prop:idset}
Under Assumption~\ref{ass:aux}, $L_n(\theta,\beta_n)=0$ for each $n\geq1$ and each $\theta\in\Theta_I$, almost surely with respect to the distribution of $(X_i)_{i=1}^\infty$.
\end{proposition}

Proposition~\ref{prop:idset} ensures validity, not sharpness. The zero set of the binding function at the pseudo-true value, $\Theta_n(s):=\{\theta\in\Theta:\;L_n(\theta,\beta_n)=0\}$, is in general a strict superset of the sharp identified region $\Theta_I$. It collects the parameter values for which model compatible data generating processes can reproduce the features of the data captured by the estimating equations~$s$. A test of $L_n(\theta,\beta_n)=0$ has no power to reject elements of $\Theta_n(s)\setminus\Theta_I$. Enriching the auxiliary model, i.e., adding estimating equations, weakly shrinks $\Theta_n(s)$ toward $\Theta_I$. In the limiting case where the estimating equations characterize the conditional distribution $P_{Y\vert X,0}$ on the relevant support, the two sets coincide.  

In practice, as in conventional indirect inference for complete models, auxiliary-model choice could draw on researchers’ understanding of which variation in the data is informative about the structural parameters. Economic reasoning and knowledge of the application can guide the selection of outcome statistics, covariates, and interactions that capture this variation. A well-chosen auxiliary model can thus target informative features of the data while keeping estimation and computation tractable. The following examples explain which information particular auxiliary models retain. Our Monte Carlo experiments in Section \ref{sec:simulation} further illustrate how auxiliary-model choice affects the informativeness of inference.

\subsubsection*{Examples: auxiliary models}
Auxiliary-model choice should balance tractability and informativeness. The pseudo-true value~$\beta_n$ must be well defined and readily estimable, and the distribution of its estimator~$\hat\beta_n$ must admit a simulation or resampling approximation, as in Assumptions~\ref{ass:aux}--\ref{ass:boot}. The auxiliary estimating equations should also capture data variation informative about the structural parameters. The following examples illustrate auxiliary-model construction and estimation, and clarify which features of the data the estimating equations retain. We discuss computational tractability of the binding function after introducing its dual representation in Proposition~\ref{prop:dual}.

\begin{continued}[Example~\ref{ex:entry} continued:]
\label{ex:aux-entry}
Consider the two-player entry game of \cite{BR:91} and \cite{Tamer:2003}. The endogenous outcome is the pair of entry indicators~$Y=(Y^1,Y^2)\in\mathcal Y=\{0,1\}^2$, and firm~$j$ has entry profit~$x^\top\delta_j+\Delta_j Y^{-j}+U^j$, with~$\Delta_j\leq0$, $j=1,2$, and zero profit from nonentry. A pure strategy equilibrium requires entry profit to be nonnegative when $Y^j=1$ and nonpositive when $Y^j=0$; either action is permitted at zero profit. The correspondence~$\Gamma(u\vert x;\theta_1)$, with~$\theta_1=(\delta_1,\delta_2,\Delta_1,\Delta_2)$, collects the pure strategy Nash equilibria of the game, and is multi-valued on the region of realizations of~$(u,x)$ where both monopoly outcomes~$(1,0)$ and~$(0,1)$ are equilibria. The distribution~$Q_{U\vert x;\theta_2}$ of~$U=(U^1,U^2)$ is bivariate normal with zero means, unit marginal variances and correlation~$\theta_2=\rho$.

Following the empirical strategy of \cite{Collard-Wexler:2013}, take as auxiliary model a seemingly unrelated linear probability model for the outcome indicators. Let~$d(y):=(y^1,y^2,y^1y^2)^\top$ and let~$w(x)\in\mathbb R^k$ be a vector of transformations of the covariates including a constant. The auxiliary score is
\begin{eqnarray*}
s(y\vert x;\beta) & := & \mbox{vec}\left[ \left( d(y)-Bw(x) \right) w(x)^\top \right], \qquad \beta:=\mbox{vec}(B)\in\mathbb R^{3k}.
\end{eqnarray*}
Equation~(\ref{eq:aux}) is then the system of empirical normal equations, whose unique solution is
\begin{eqnarray*}
B_n & = & \left( \sum_{i=1}^n \mathbb E\left[ d(Y)\vert X_i\right] w(X_i)^\top \right) \left( \sum_{i=1}^n w(X_i)w(X_i)^\top \right)^{-1},
\end{eqnarray*}
and $\beta_n = \mbox{vec}(B_n)$.
The estimator~$\hat\beta_n$ is the ordinary least squares regression of~$d(Y_i)$ on~$w(X_i)$.

In this example, the conditional means of~$Y^1$, $Y^2$, and~$Y^1Y^2$ determine the entire conditional distribution of entry outcomes, which is the observable information used to define the sharp identified region. The auxiliary model provides a reduced-form approximation to these conditional mean functions by projecting them onto the function space spanned by the components of~$w(x)$ under the empirical covariate distribution~$P_{X,n}$. The information reduction comes from retaining only these projections. Adding covariate transformations allows the auxiliary model to capture additional variation in the conditional entry probabilities, further closing the gap between $\Theta_n(s)$ and $\Theta_I$. Alternatively, given the normally distributed payoff shocks, one could also use a probit regression as a potentially misspecified auxiliary model for a firm's entry decision instead of a linear regression model.

\end{continued}

\begin{continued}[Example~\ref{ex:auctions} continued:]
\label{ex:aux-auction}
Consider the incomplete model of English auctions of \cite{HT:2003} (see \cite{CR:2017} for a version with unobserved heterogeneity). An auction with~$m\geq2$ bidders with nonnegative valuations yields the vector~$Y=(B^1,\ldots,B^m)$ of final bids, the covariates~$X$ collect auction characteristics, including the number of bidders~$m$ and the minimum bid increment~$\delta$, and the unobservable~$U=(V^1,\ldots,V^m)$ collects the private valuations, independent and identically distributed conditionally on~$x$ with distribution parameterized by~$\theta_2$. The model does not specify bidding behavior. It imposes only the two behavioral axioms that no bidder bids more than their valuation, and that no bidder allows an opponent to win at a price they are willing to beat. Writing~$v_{(j:m)}$ and~$b_{(j:m)}$ for the order statistics of valuations and bids, the correspondence is the union of polyhedra indexed by the winning bidder $j^\ast$. A winner must have a highest bid, and every losing bidder must have valuation at most the winning price plus the bid increment $\delta\geq0$. At equality a losing bidder may decline to bid, and tied highest bids may be awarded to any of the tied bidders. Thus
\begin{eqnarray*}
\Gamma(v\vert x;\theta_1) & = & \bigcup_{j^\ast=1}^m \left\{ b\in\mathbb R_+^m:\;
\begin{array}{ll}
b^j\leq v^j, & j=1,\ldots,m,\\
b^j\leq b^{j^\ast},\quad v^j\leq b^{j^\ast}+\delta, & j\neq j^\ast
\end{array}\right\},
\end{eqnarray*}
which is multi-valued because bids may lie anywhere between the implied bounds; the model is therefore incomplete in the sense of Section~\ref{sec:model}.

The auxiliary model is a seemingly unrelated regression of the top order statistics of the bids on the auction covariates. Let~$t(y)\coloneqq(b_{(m:m)},b_{(m-1:m)})^\top$, the transaction price and the highest losing bid, and let~$w(x)\in\mathbb R^k$ be a vector of transformations of the covariates, including a constant and functions of the number of bidders~$m$. The auxiliary score is
\begin{eqnarray*}
s(y\vert x;\beta) & := & \mbox{vec}\left[ \left( t(y)-Bw(x) \right) w(x)^\top \right], \qquad \beta:=\mbox{vec}(B)\in\mathbb R^{2k},
\end{eqnarray*}
so that the pseudo-true value~$\beta_n$ is the vector of empirical projection coefficients, and the estimator~$\hat\beta_n$ is ordinary least squares.

The two highest bids are natural targets for the auxiliary model because the behavioral restrictions bound the second-highest valuation between the highest losing bid and the winning bid plus the minimum bid increment. The auxiliary model retains projections of their conditional mean functions onto the space spanned by the components of~$w(x)$. It therefore captures selected identifying variation while omitting other features of their joint conditional distribution. Lower bids can also provide some identifying information because each bounds its bidder's valuation from below. Including statistics of lower bids can therefore help the auxiliary model retain additional identifying information.
\end{continued}

\begin{continued}[Example~\ref{ex:networks} continued:]
\label{ex:aux-network}

The data consist of a sample of $n$ independent networks (villages, classrooms, markets),
where network $i$ has $m_i$ members. The endogenous outcome for network $i$ is
$Y_i=(G_i,y_i)$, where $G_i$ is the adjacency matrix of the network and
$y_i\in\mathbb{R}^{m_i}$ is the vector of member outcomes. The exogenous covariates are
$X_i=(m_i,x_i)$, with $x_i$ the matrix of member characteristics. The structural model
specifies preferences over links and a nonlinear outcome equation with peer effects, and
predicts the pairs $(g,y)$ such that $g$ is pairwise stable given $(x,u;\theta_1)$ and $y$
solves the system of best responses on $g$. Multiplicity of pairwise stable networks and,
possibly, of outcome equilibria makes the correspondence
$\Gamma(u\mid x;\theta_1)$ multi-valued. Neither linearity of peer effects nor exogeneity
of the network is imposed.

Some classical auxiliary models can approximate the outcome and
network-formation aspects of the data generating process. For the outcome generating process, consider the linear-in-means model of \citet{Manski:93}, which treats the observed network as
exogenous. Let $\widetilde G$ denote the row-normalized adjacency matrix and $\iota$ the
vector of ones, and consider
\[
y=\alpha\iota+\gamma\widetilde G y+x\delta+\widetilde Gx\eta+\varepsilon,
\]
estimated by instrumental variables with the instruments of \citet{BDF:2009},
based on characteristics of second-degree neighbors. Choosing a just-identified
instrument set,
\[
Z(Y,X):=[\iota,\widetilde G^{\,2}x^{(1)},x,\widetilde Gx]
\]
of the same dimension as the regressor matrix
\[
W(Y,X):=[\iota,\widetilde Gy,x,\widetilde Gx],
\]
the corresponding auxiliary score is
\[
s^{y}(Y\mid X;\beta_y)
=
Z(Y,X)^\top\bigl(y-W(Y,X)\beta_y\bigr),
\qquad
\beta_y=(\alpha,\gamma,\delta^\top,\eta^\top)^\top .
\]
Since $s^y$ is affine in $\beta_y$, the analogue of \eqref{eq:aux} has the unique solution
\[
\beta_{y,n}
=
\left(
\sum_{i=1}^n E[Z^\top W\mid X_i]
\right)^{-1}
\sum_{i=1}^n E[Z^\top y\mid X_i],
\]
whenever the aggregated matrix in parentheses is nonsingular, a population analogue of
the rank condition of \citet{BDF:2009} evaluated under the true data generating
process. The fact that the instruments are constructed from the endogenous network $G$
is immaterial: Assumption~\ref{ass:aux} restricts only the functional form of the auxiliary estimating
equations, not their structural interpretation, and validity of the confidence region does
not require the auxiliary orthogonality conditions to be correctly specified.

The network-formation part of the data generating process can be approximated using auxiliary models from the empirical networks literature. See Section~13.4 of \citet{Chandrasekhar:2016}. One simple choice is a vector $S(G)\in\mathbb{R}^{d_g}$ of normalized subgraph counts---links, triangles, $k$-stars, and possibly interactions with node characteristics---and the estimating equation
\[
s^{g}(Y\mid X;\beta_g):=S(G)-\beta_g.
\]
Then
\[
\beta_{g,n}
=
\frac{1}{n}\sum_{i=1}^n E[S(G)\mid X_i]
\]
exists and is unique by construction, and its sample analogue is a sample average across
networks.

Alternatively, one may use an exponential random graph model estimated by maximum
pseudo-likelihood. If $t_{jk}(G)$ denotes the vector of change statistics associated with
link $(j,k)$ and $\Lambda$ the logistic distribution function, the corresponding score is
\[
s^{g}(Y\mid X;\beta_g)
=
\sum_{j<k}
\bigl(
G_{jk}-\Lambda(\beta_g^\top t_{jk}(G))
\bigr)t_{jk}(G).
\]
The expected pseudo-log-likelihood is concave in $\beta_g$, and strictly concave when the
expected design matrix is positive definite.

These auxiliary components may be used separately or stacked,
\[
s(Y\mid X;\beta)
=
\begin{pmatrix}
s^{y}(Y\mid X;\beta_y)\\
s^{g}(Y\mid X;\beta_g)
\end{pmatrix},
\qquad
\beta=(\beta_y^\top,\beta_g^\top)^\top,
\]
thereby allowing the auxiliary model to capture both peer-outcome behavior and salient
features of the endogenous network. 

These auxiliary models build on familiar reduced-form empirical specifications for peer-outcome relationships and network formation. They summarize economically interpretable features of the data, including associations between outcomes and neighbors’ characteristics, network density, and clustering. To the extent that these reduced-form features reflect the structural mechanisms of interest, the auxiliary models can retain useful identifying information. Additional outcome equations or network statistics can be included to capture further relevant variation.
\end{continued}

Other possible auxiliary models include specific equilibrium selections in models with multiple equilibria, linearized DSGE models with Gaussian errors and the reduced form of structural vector auto-regressions.

\subsection{Dual characterization of the binding function}
\label{sec:duality}

For the computational characterization and the subsequent analysis of the binding function, we impose the following regularity conditions on the structural correspondence and on the auxiliary estimating function.

\begin{assumption}[Regularity]
\label{ass:reg}
For each $\theta=(\theta_1,\theta_2)\in\Theta$:
\begin{enumerate}[label=(\alph*)]
\item The spaces $\mathcal Y$ and $\mathcal U$ are Polish, and $x\mapsto Q_{U\vert x;\theta_2}$ is a Markov kernel from $\supp(X)$ to $\mathcal U$. The correspondence $(u,x)\mapsto\Gamma(u\vert x;\theta_1)$ from $\mathcal U\times\supp(X)$ to $\mathcal Y$ has nonempty compact values and is \emph{jointly} measurable, in the sense that
\begin{eqnarray*}
&&\left\{ (u,x)\in\mathcal U\times\supp(X):\;\Gamma(u\vert x;\theta_1)\cap G\neq\varnothing \right\} \mbox{ is measurable for all open } G\subseteq\mathcal Y.
\end{eqnarray*}
\item The set $\mathbb B_0$ is nonempty, closed and convex, and for every $x\in\supp(X)$, the map $(y,\beta)\mapsto s(y\vert x;\beta)$ is continuous on $\mathcal Y\times\mathbb B_0$ and continuously differentiable in $\beta$, with jointly continuous derivative $(y,\beta)\mapsto\nabla_\beta s(y\vert x;\beta)$. In addition, $(y,x)\mapsto s(y\vert x;\beta)$ is jointly measurable for each $\beta\in\mathbb B_0$.
\item The function
\begin{eqnarray*}
b(u,x;\theta) & := & \sup_{\beta\in\mathbb B_0}\;\sup_{y\in\Gamma(u\vert x;\theta_1)}\;\max\left\{ \left\| s(y\vert x;\beta)\right\|,\; \left\|\nabla_\beta s(y\vert x;\beta) \right\| \right\}
\end{eqnarray*}
satisfies 
\begin{eqnarray*}
    \forall x\in\supp(X),\;\mathbb E_{Q_{U\vert x;\theta_2}}\left[ b(U,x;\theta)\right]<\infty, \mbox{ and }
    \mathbb E_{P_{X,0}}\mathbb E_{Q_{U\vert X;\theta_2}}\left[ b(U,X;\theta)\right]<\infty.
\end{eqnarray*}
\item The set $\mathbb B_0$ has nonempty interior, and there exists $\delta_\ast>0$ such that, almost surely with respect to the distribution of $(X_i)_{i=1}^\infty$, $\{\beta\in\real^{d_\beta}:\;\|\beta-\beta_n\|\leq\delta_\ast\}\subseteq\mathbb B_0$ for all $n$ large enough.
\end{enumerate}
\end{assumption}

Given part~(a), the structural model~$\mathcal P_\theta$ is automatically non-empty: the correspondence $(u,x)\mapsto\Gamma(u\vert x;\theta_1)$ is measurable with nonempty closed values in a Polish space, so the Kuratowski--Ryll-Nardzewski selection theorem \citep[Theorem~18.13]{AB:99} provides a jointly measurable selection $\gamma$ with $\gamma(u,x)\in\Gamma(u\vert x;\theta_1)$ for all $(u,x)$; the pushforward $\pi^0_x$ of $Q_{U\vert x;\theta_2}$ under $u\mapsto(\gamma(u,x),u)$ is then a Markov kernel satisfying the support and marginal restrictions of Definition~\ref{def:DGP}, so that its $\mathcal Y$-marginal belongs to $\mathcal P_\theta$. 
In the leading example of Section~\ref{sec:model}, where $\mathcal Y$ is finite and $\Gamma$ is defined by finitely many inequalities in $(u,x,\theta_1)$, joint measurability is immediate and nonemptiness amounts to existence of a pure strategy equilibrium, of a rationalizable bid vector, or of a pairwise stable network. Part~(c) is satisfied for each fixed~$x$ when $s$ and $\nabla_\beta s$ are continuous and~$\mathcal Y\times \cl(\mathbb B_0)$ is compact. Part~(d) holds for instance if $\beta_n$ converges along almost every covariate sequence to a point in the interior of $\mathbb B_0$, with a deterministic positive lower bound on the distance of that limit to the boundary.

We derive a computationally feasible dual characterization of the binding function. For $u\in\mathcal U$ and $1\leq i\leq n$, let $\Gamma^s_i(u):=s(\Gamma(u\vert X_i;\theta_1)\vert X_i;\beta)\subseteq\real^{d_s}$ denote the image of the prediction set under the auxiliary estimating function, and let ${\bf\Gamma}^s_i$ denote the random closed set induced by the correspondence $\Gamma^s_i$ and the distribution $Q_{U\vert X_i;\theta_2}$.

\begin{proposition}[Duality]
\label{prop:dual}
Under Assumption~\ref{ass:reg}, for each $n\geq1$ and all $(\theta,\beta)\in\Theta\times\mathbb B_0$,
\begin{eqnarray}
\label{eq:dual}
L_n(\theta,\beta) & = & \sup_{\lambda\in\mathcal S} \left( \frac1n\sum_{i=1}^n \mathbb E_{Q_{U\vert X_i;\theta_2}} \left[ \inf_{y_i\in\Gamma(U\vert X_i;\theta_1)} \lambda^\top s \left( y_i\vert X_i;\beta \right) \vert X_i \right] \right)
\end{eqnarray}
for all realizations $(X_1,\ldots,X_n)$, where $\mathcal S$ is the closed unit ball in $\real^{d_s}$.
\end{proposition}

The dual representation~(\ref{eq:dual}) replaces the infinite dimensional program of Definition~\ref{def:bind-n} with a finite dimensional maximization problem over the unit ball of $\real^{d_s}$. The dual objective is concave and positively homogeneous of degree one in $\lambda$, as an average of expectations of infima of linear functions of $\lambda$, so that the outer problem is a well behaved concave program. The inner problem, the minimization of the scalar function $y\mapsto\lambda^\top s(y\vert X_i;\beta)$ over the prediction set $\Gamma(U\vert X_i;\theta_1)$, is a simple enumeration when the prediction set is sufficiently small, as in Example~\ref{ex:entry}. Finally, the conditional expectations $\mathbb E_{Q_{U\vert X_i;\theta_2}}[\,\cdot\,]$ can be approximated by Monte Carlo integration over simulated draws of the latent variables, so that, in keeping with the spirit of indirect inference, the procedure only requires the ability to simulate from the latent variable distribution and to compute the prediction correspondence.

The proof of Proposition~\ref{prop:dual} shows that $L_n(\theta,\beta)$ is the distance from the origin to the compact convex set 
\begin{eqnarray*}
    \mathcal K_n(\theta,\beta) & := & \frac1n\bigoplus_{i=1}^n{\bf E}{\bf\Gamma}^s_i,
\end{eqnarray*} 
which is the Minkowski average of the convexified selection expectations of the random sets~${\bf\Gamma}^s_i$. The set $\mathcal K_n(\theta,\beta)$ collects the values of the averaged auxiliary estimating equations attainable by model compatible data generating processes. In particular, $L_n(\theta,\beta_n)=0$ if and only if $0\in \mathcal K_n(\theta,\beta_n)$, and the geometry of $\mathcal K_n$ near the origin, specifically whether $0$ lies in the interior or on the boundary of $\mathcal K_n(\theta,\beta_n)$, governs the local behavior of the binding function.

\subsubsection*{Examples: computation of the binding function}

The second tractability requirement concerns the inner minimization~$\inf_{y\in\Gamma(U\vert X;\theta_1)}\lambda^\top s(y\vert X;\beta)$ in~(\ref{eq:dual}). This minimization can be evaluated by enumeration when the prediction set is sufficiently small, or by exploiting the structure of the auxiliary score and the structural correspondence.

\begin{continued}[Example~\ref{ex:entry} continued:]
There are only four possible entry profiles in this two-player game.  Thus, the inner minimization in~(\ref{eq:dual}) can be computed by enumeration of the feasible entry profiles, and~$y\mapsto\lambda^\top s(y\vert x;\beta)$ is an affine function of~$d(y)$. 
\end{continued}

\begin{continued}[Example~\ref{ex:auctions} continued:]
Recall the auxiliary model is a sieve-type prediction for $t(y) \coloneqq(b_{(m:m)},b_{(m-1:m)})^\top$.
For the dual characterization~(\ref{eq:dual}), the map~$y\mapsto\lambda^\top s(y\vert x;\beta)$ is affine in~$t(y)$, and the axioms of the model constrain~$t(y)$ to a two-dimensional polytope described by the order statistics of the valuations: $b_{(m:m)}\in[\max\{0,v_{(m-1:m)}-\delta\},\,v_{(m:m)}]$ and~$0\leq b_{(m-1:m)}\leq\min\{v_{(m-1:m)},\,b_{(m:m)}\}$, together with any reserve price constraint. 
The inner minimization is therefore a two-dimensional linear program, solved in closed form at a vertex of the polytope.
\end{continued}

\begin{continued}[Example~\ref{ex:networks} continued:]
    In this example, the inner minimization in \eqref{eq:dual} is over pairwise stable networks and their associated outcome equilibria. A bounded-degree restriction, as in \cite{dPRST:2018}, limits the set of candidate networks and can help reduce the computational burden. When the remaining networks can be solved efficiently and optimization over their associated outcome equilibria is tractable, the inner minimum can be computed by comparing the minimized criterion values across all pairwise stable networks.
\end{continued}

\subsection{Lipschitz continuity and directional differentiability}

Finally, using characterization~(\ref{eq:dual}) of the binding function, we show that the binding function is Lipschitz and Hadamard directionally differentiable (see \citealp{Shapiro:90} for a definition) at $\beta_n$, and we derive an explicit expression for the directional derivative. For each $\theta\in\Theta$, $\beta\in\mathbb B_0$ and $\lambda\in\mathcal S$, write
\begin{eqnarray}
\label{eq:phin}
\varphi_n(\lambda,\beta) & := & \frac1n\sum_{i=1}^n \mathbb E_{Q_{U\vert X_i;\theta_2}}\left[ \inf_{y\in\Gamma(U\vert X_i;\theta_1)} \lambda^\top s(y\vert X_i;\beta)\;\Big\vert\; X_i\right]
\end{eqnarray}
for the dual objective, so that $L_n(\theta,\beta)=\sup_{\lambda\in\mathcal S}\varphi_n(\lambda,\beta)$ by Proposition~\ref{prop:dual}, and define the solution sets of the inner and outer programs:
\begin{eqnarray*}
\hat{\mathcal Y}_i(u,\lambda,\beta) \; := \; \argmin_{y\in\Gamma(u\vert X_i;\theta_1)}\lambda^\top s(y\vert X_i;\beta),
& \quad &
\hat\Lambda_n(\theta,\beta) \; := \; \argmax_{\lambda\in\mathcal S}\;\varphi_n(\lambda,\beta).
\end{eqnarray*}
Under Assumption~\ref{ass:reg}, both solution sets are nonempty and compact. The directional differentiability of the binding function requires one additional regularity condition on the inner program.

\begin{assumption}[Common projected gradients at inner minimizers]
\label{ass:uniq}
For each $\theta\in\Theta$ and each $n\geq1$, almost surely with respect to the distribution of $(X_i)_{i=1}^\infty$: for every $\lambda\in\hat\Lambda_n(\theta,\beta_n)$ and each $1\leq i\leq n$, the set $\{\lambda^\top\nabla_\beta s(y\vert X_i;\beta_n): y\in \hat{\mathcal Y}_i(u,\lambda,\beta_n)\}$ is a singleton for $Q_{U\vert X_i;\theta_2}$-almost every $u$.
\end{assumption}

Assumption~\ref{ass:uniq} requires all inner minimizers to yield the same gradient of $\lambda^\top s(y\vert X_i;\beta)$ with respect to $\beta$ at $\beta_n$. The condition is imposed only at the pseudo-true value and for maximizing directions $\lambda\in\hat\Lambda_n(\theta,\beta_n)$. A sufficient condition is that $\hat{\mathcal Y}_i(u,\lambda,\beta_n)$ be a singleton for each nonzero maximizing direction. The assumption also holds whenever $\nabla_\beta s(y\vert x;\beta)$ does not depend on $y$. This includes least-squares auxiliary regressions of outcomes on covariates, such as those used in the entry and auction examples.

\begin{proposition}[Lipschitz continuity and directional differentiability]
\label{prop:hdd}
Let Assumptions~\ref{ass:aux}, \ref{ass:reg} and~\ref{ass:uniq} hold. Then, for each $\theta\in\Theta$ and each $n\geq1$, almost surely with respect to the distribution of $(X_i)_{i=1}^\infty$:
\begin{enumerate}[label=(\roman*)]
\item $\beta\mapsto L_n(\theta,\beta)$ is Lipschitz on $\mathbb B_0$, with Lipschitz constant 
\begin{eqnarray*}
    \ell_n & := & \frac1n\sum_{i=1}^n\mathbb E_{Q_{U\vert X_i;\theta_2}}[b(U,X_i;\theta)]
    \quad \mbox{ and } \quad \sup_{n\geq1}\ell_n \; < \; \infty;
\end{eqnarray*} 
\item For all sufficiently large $n$, $\beta\mapsto L_n(\theta,\beta)$ is Hadamard directionally differentiable at the pseudo-true value $\beta_n$ of~(\ref{eq:aux}), with directional derivative given for every direction $d\in\real^{d_\beta}$ by:
\begin{eqnarray}
\label{eq:hdd}
L^\prime_{\beta_n}(\theta,d) & = & \max_{\lambda\in\hat\Lambda_n(\theta,\beta_n)}\;\frac1n\sum_{i=1}^n \mathbb E_{Q_{U\vert X_i;\theta_2}}\left[ \min_{y\in\hat{\mathcal Y}_i(U,\lambda,\beta_n)} \lambda^\top \nabla_\beta s(y\vert X_i;\beta_n)\, d \;\Big\vert\; X_i\right].
\end{eqnarray}
\end{enumerate}
\end{proposition}

The inference procedure of Section~\ref{sec:inference} evaluates the binding function at perturbed arguments of the form $\hat\beta_n+\epsilon_nh$, where $h$ ranges over the support of the resampling distribution and is therefore unbounded in the leading Gaussian case. No fixed neighborhood of $\beta_n$ contains all such points, so Assumption~\ref{ass:reg}(d) alone does not make these evaluations well defined. We therefore extend the binding function to the whole of $\real^{d_\beta}$.

\begin{definition}[Projection extension of the binding function]
\label{def:ext}
For each $\theta\in\Theta$ and $n\geq1$, define, for all $\beta\in\real^{d_\beta}$,
$\tilde L_n(\theta,\beta):=L_n(\theta,\Pi_{\mathbb B_0}(\beta)),$
where $\Pi_{\mathbb B_0}$ is the Euclidean projection operator on~$\mathbb B_0$, which is well defined and $1$-Lipschitz because $\mathbb B_0$ is nonempty, closed and convex. From Section~\ref{sec:inference} onwards, and with a slight abuse of notation, $L_n(\theta,\cdot)$ denotes $\tilde L_n(\theta,\cdot)$.
\end{definition}

The map $\tilde L_n(\theta,\cdot)$ is an extension of $L_n(\theta,\cdot)$. First, it agrees with $L_n(\theta,\cdot)$ on $\mathbb B_0$. Consequently Definition~\ref{def:bind-n}, Propositions~\ref{prop:idset} and~\ref{prop:dual}, the identity $L_n(\theta,\beta_n)=0$ on $\Theta_I$, and the confidence region itself are unaffected, and the extension is invisible whenever the perturbed argument happens to lie in $\mathbb B_0$, which by Assumption~\ref{ass:reg}(d) is an event of probability tending to one for the perturbations that matter. Second, $\tilde L_n(\theta,\cdot)$ is $\ell_n$-Lipschitz on all of $\real^{d_\beta}$, which is what Step~2 of the proof of Theorem~\ref{thm:ndm} requires uniformly in the perturbation direction. Third, $\tilde L_n\geq0$, since $L_n\geq0$, which is used in the degenerate case of that proof. For all sufficiently large $n$, $\tilde L_n$ coincides with $L_n$ on the ball of radius $\delta_\ast$ around $\beta_n$ and is therefore Hadamard directionally differentiable at $\beta_n$ with the same derivative~(\ref{eq:hdd}), so Proposition~\ref{prop:hdd} carries over to the extension.

\section{Inference on the structural parameter}
\label{sec:inference}

The objective of this section is to propose a confidence region that covers any parameter value in the identified set $\Theta_I$ with probability asymptotically at least $1-\alpha$, for some prescribed $\alpha\in(0,1)$. Since the binding function is only directionally differentiable, standard delta method arguments and the bootstrap fail (see \citealp{FS:2019}). We rely instead on the numerical delta method proposed by \cite{HL:2018}.

\begin{definition}[Numerical directional derivative]
\label{def:ndd}
For any $\theta\in\Theta$, $h\in\real^{d_\beta}$, and any sequence $\epsilon_n$ satisfying $\epsilon_n\downarrow0$ and $r_n\epsilon_n\rightarrow\infty$, define, with $L_n$ the extension of Definition~\ref{def:ext},
\begin{eqnarray*}
\hat L^\prime_{\beta_n}(\theta,h) & := & \frac{1}{\epsilon_n}\left( L_n(\theta,\hat\beta_n+\epsilon_nh)-L_n(\theta,\hat\beta_n) \right).
\end{eqnarray*}
\end{definition}

For any $\tau\in(0,1)$, call $\hat c_{n,\tau}(\theta)$ the $\tau$-quantile of the distribution of $\hat L^\prime_{\beta_n}(\theta,\mathbb Z_n^\ast)$, conditional on the data.
The conditions $\epsilon_n\downarrow0$ and $r_n\epsilon_n\rightarrow\infty$ ensure that the numerical differencing step is small enough for the difference quotient to approximate the directional derivative, yet large relative to the estimation error in $\hat\beta_n$, so that replacing $\beta_n$ with $\hat\beta_n$ in the base point is innocuous; see the proof of Theorem~\ref{thm:ndm}. A natural rule of thumb, which balances the two requirements, is $\epsilon_n\propto r_n^{-1/2}$, i.e., $\epsilon_n\propto n^{-1/4}$ in the standard case $r_n=\sqrt n$. \cite{HL:2018} discuss the choice of the step size.

Two further conditions are needed for the asymptotic validity of the numerical delta method in the present conditional, triangular array setting. The first strengthens the directional expansion of Proposition~\ref{prop:hdd} to hold uniformly along the covariate sequence.

\begin{assumption}\label{ass:unif}
For each $\theta\in\Theta_I$, every $c>0$ and every sequence $t_n\downarrow0$, almost surely with respect to the distribution of $(X_i)_{i=1}^\infty$,
\begin{eqnarray*}
\sup_{\|h\|\leq c}\left| \frac{L_n(\theta,\beta_n+t_nh)-L_n(\theta,\beta_n)}{t_n} - L^\prime_{\beta_n}(\theta,h) \right| & \rightarrow & 0.
\end{eqnarray*}
Note that for each fixed $c$ the perturbations are eventually feasible, $\beta_n+t_nh\in\mathbb B_0$ for all $\|h\|\leq c$ as soon as $t_nc\leq\delta_\ast$, by Assumption~\ref{ass:reg}(d), so that the display involves the binding function itself and not merely its extension.
\end{assumption}

For each sufficiently large fixed $n$, convergence of the corresponding difference quotients as the step tends to zero is the Hadamard directional differentiability established in Proposition~\ref{prop:hdd}. Because the binding function $L_n$ and the pseudo-true value $\beta_n$ vary with the sample size through the conditioning covariates, the delta method must be applied along the covariate sequence, and Assumption~\ref{ass:unif} requires the expansion to hold uniformly in $n$. It is automatically satisfied in the absence of covariates, where $L_n$ and $\beta_n$ do not depend on $n$. 

The second condition regulates the behavior of the distribution of the directional derivative near the critical value; conditions of this type are standard in bootstrap quantile arguments, see \cite{HL:2018} and \cite{FS:2019}.

\begin{assumption}\label{ass:cv}
Fix $\alpha\in(0,1)$. For each $\theta\in\Theta_I$, almost surely with respect to the distribution of $(X_i)_{i=1}^\infty$, one of the following two conditions holds, where $F_n(\cdot)$ denotes the conditional distribution function of $L^\prime_{\beta_n}(\theta,\mathbb Z_0)$ given $(X_i)_{i=1}^n$ and $c_{n,1-\alpha}(\theta)$ its $(1-\alpha)$-quantile:
\begin{enumerate}[label=(\alph*)]
\item (degenerate case) there exist $n_0\geq1$ and $\delta_0\in(0,\delta_\ast]$ such that $L_n(\theta,\beta)=0$ for all $n\geq n_0$ and all $\beta$ with $\|\beta-\beta_n\|\leq\delta_0$, a set of parameter values contained in $\mathbb B_0$ by Assumption~\ref{ass:reg}(d); or
\item (nondegenerate case) the distribution functions $F_n$ are asymptotically continuous and strictly increasing at their $(1-\alpha)$-quantiles, i.e.,
\begin{eqnarray*}
\lim_{\delta\downarrow0}\;\limsup_n\;\left[ F_n\big(c_{n,1-\alpha}(\theta)\big) - F_n\big(c_{n,1-\alpha}(\theta)-\delta\big)\right] & = & 0,
\end{eqnarray*}
and, for every sufficiently small $\delta>0$,
\begin{eqnarray*}
\liminf_n\;\left[ (1-\alpha) - F_n\big(c_{n,1-\alpha}(\theta)-\delta\big)\right] & > & 0.
\end{eqnarray*}
\end{enumerate}
\end{assumption}

Case~(a) covers parameter values in the interior of the auxiliary outer region: when the origin lies well inside the set $\mathcal K_n(\theta,\beta_n)$, the binding function vanishes on a neighborhood of $\beta_n$, the directional derivative is identically zero, and its distribution is a unit mass at zero, at which the continuity requirement of case~(b) necessarily fails; coverage at such points is nonetheless immediate, as the proof shows. Case~(b) covers the binding boundary points, where the test has nontrivial local behavior, and requires that no mass of the limiting distribution of the test statistic accumulate at the critical value, nor that its distribution function be flat immediately below it. When the conditional law of $L^\prime_{\beta_n}(\theta,\mathbb Z_0)$ stabilizes as $n\rightarrow\infty$, case~(b) reduces to the familiar requirement that the limiting distribution function be continuous and strictly increasing at its $(1-\alpha)$-quantile. This requirement depends on the chosen $\alpha$: for example, if the derivative law is that of $\max\{0,Z\}$ with $Z$ a centered nondegenerate Gaussian variable, its atom at zero rules out case~(b) when $\alpha\geq1/2$. The coverage results below apply at confidence levels satisfying Assumption~\ref{ass:cv}.

\begin{theorem}
\label{thm:ndm}
Fix $\alpha\in(0,1)$ for which Assumption~\ref{ass:cv} holds. Let Assumptions~\ref{ass:aux}--\ref{ass:unif} hold, and adopt the notation of Definitions~\ref{def:DGP}--\ref{def:ndd}. Then, for each $\theta\in\Theta_I$,
\begin{eqnarray*}
\limsup_n\;\mathbb P\left(r_nL_n(\theta,\hat\beta_n)>\hat c_{n,1-\alpha}(\theta) \right) & \leq & \alpha.
\end{eqnarray*}
\end{theorem}

The confidence region for the structural parameter is then
\begin{eqnarray*}
\mathcal C_{n,1-\alpha} & := & \left\{ \theta\in\Theta:\; r_nL_n(\theta,\hat\beta_n)\leq\hat c_{n,1-\alpha}(\theta) \right\},
\end{eqnarray*}
and Theorem~\ref{thm:ndm} states that $\mathcal C_{n,1-\alpha}$ covers each point of the sharp identified region with asymptotic probability at least $1-\alpha$. The probability is with respect to the joint distribution of the data and of the resampling scheme. Note that the test statistic is scaled by the rate $r_n$ of the auxiliary estimator: under the null hypothesis $\theta\in\Theta_I$, the binding function vanishes at $\beta_n$, so that $L_n(\theta,\hat\beta_n)$ is of order $r_n^{-1}$ and $r_nL_n(\theta,\hat\beta_n)$ is the correctly centered and scaled statistic whose distribution the numerical delta method approximates.

The confidence region is computed as follows. First, estimate the auxiliary parameter $\hat\beta_n$ from the data, and generate draws $\mathbb Z^\ast_{n,1},\ldots,\mathbb Z^\ast_{n,B}$ by simulation from the estimated limit distribution or by a resampling scheme that approximates the conditional estimator law, such as the construction in Appendix~\ref{sec:cross-fit}. Then, for each candidate value $\theta$ on a grid or along an adaptive search: (i) compute $L_n(\theta,\hat\beta_n)$ from the dual representation~(\ref{eq:dual}); (ii) compute the numerical directional derivatives $\hat L^\prime_{\beta_n}(\theta,\mathbb Z^\ast_{n,b})$, $b=1,\ldots,B$, each of which requires one additional evaluation of the binding function at $\hat\beta_n+\epsilon_n\mathbb Z^\ast_{n,b}$; (iii) take $\hat c_{n,1-\alpha}(\theta)$ to be the empirical $(1-\alpha)$-quantile of these $B$ draws; and (iv) include $\theta$ in $\mathcal C_{n,1-\alpha}$ if and only if $r_nL_n(\theta,\hat\beta_n)\leq\hat c_{n,1-\alpha}(\theta)$. The cost per candidate $\theta$ is thus $B+1$ evaluations of a finite dimensional concave program.

As noted in Section~\ref{sec:sharp}, the sharp identified region may be empty, in which case the structural model is incompatible with the data. Since $\mathcal C_{n,1-\alpha}$ contains each point of $\Theta_I$ with asymptotic probability at least $1-\alpha$, an empty confidence region, which occurs whenever~$\inf_{\theta\in\Theta}\left(r_nL_n(\theta,\hat\beta_n)-\hat c_{n,1-\alpha}(\theta)\right)>0$, constitutes evidence against the hypothesis that the structural model is correctly specified.

Finally, the binding function of Definition~\ref{def:bind-n} depends on the scale and orientation of the auxiliary estimating equations. Replacing $s$ with $As$ for a nonsingular matrix $A$ leaves the zero set $\Theta_n(s)$, hence the loss of sharpness, unchanged, but can change $L_n$, its directional derivative and the shape of the confidence region. As in the Wald and score forms of indirect inference \citep{GMR:93,GT:96}, this dependence can be removed by weighting the estimating equations with the inverse square root of their conditional covariance matrix, i.e., by replacing $s$ with $W_ns$, where $W_n=B_{c,n}^{-1/2}$ and $B_{c,n}$ is defined in Appendix~\ref{sec:cross-fit}. This choice standardizes the estimating equations given the covariates and makes the weighted binding function invariant to nonsingular linear transformations of $s$, provided the covariance matrix is transformed accordingly. However, this conditional covariance matrix needs to be estimated in practice. Appendix~\ref{sec:weights} extends Propositions~\ref{prop:idset}--\ref{prop:hdd} and Theorem~\ref{thm:ndm} to weighted binding functions, and shows that the weights $W_n$ may be replaced by estimates $\hat W_n$ satisfying $\|\hat W_nW_n^{-1}-I_{d_s}\|\rightarrow0$ in probability, as specified in Assumption~\ref{ass:weights}. No convergence rate beyond consistency is required. This is because the estimation error of the weighting matrix perturbs the binding function multiplicatively, by a factor close to one uniformly in $(\theta,\beta)$, so that the binding function remains close to zero with $\hat W_n$ wherever it is close to zero with $W_n$, and the test statistic and critical values are asymptotically unaffected. Since $\hat W_n$ only enters the inner programs of the dual representation~(\ref{eq:dual}), the cost per candidate value of $\theta$ is unchanged.

\section{Monte Carlo evidence} \label{sec:simulation}
We use a binary entry game with up to 100 potential entrants to assess pointwise coverage, computational feasibility, and the informativeness of our indirect inference procedure. To examine the role of auxiliary-model choice, we compare two auxiliary models that differ in their ability to capture data variation informative about the structural parameters. The results illustrate how a carefully chosen auxiliary model that exploits relevant variation could produce more informative confidence intervals.

\subsection{Design and auxiliary models}

Each sample contains $n=5{,}000$ independent markets. A market has $J\in\{10,20,50,100\}$ potential entrants, comprising four big firms with indices in $G_B$ and $J-4$ small firms with indices in $G_S$. Firm $j$ earns zero from nonentry and, if it enters market $i$, earns
\begin{equation}
\label{eq:simulation-profit}
\pi_{ij}(Y_{i,-j})
=\alpha X_{ij}-\gamma_B N_{B,-ij}
-\frac{\gamma_S}{J-4}N_{S,-ij}+U_{ij},
\end{equation}
where $N_{B,-ij} = \sum_{j'\in G_B} \mathbbm{1}(j'\neq j)Y_{i,j'}$ and $N_{S,-ij} = \sum_{j'\in G_S} \mathbbm{1}(j'\neq j)Y_{i,j'}$ count the other big- and small-firm entrants. Both competition coefficients are nonnegative. The normalization of the small-firm count keeps its aggregate competitive effect comparable across specifications with different $J$. The structural correspondence contains all pure-strategy Nash equilibria.

The shocks $U_{ij}$ are independent standard normal random variables, independent of the covariates. Covariates are drawn independently across firms and markets, with $X_{ij}\sim N(2.7,1)$ for big firms and $X_{ij}\sim N(1,1)$ for small firms. For each $J$, the covariate panel is drawn once and held fixed across Monte Carlo repetitions. The true structural parameter is $\theta_0=(\alpha_0,\gamma_{B0},\gamma_{S0})=(1,1,4)$. To generate outcomes when equilibria are multiple, we select an equilibrium with the largest number of big entrants, then the largest total number of entrants, and finally the highest latent payoff indices $\alpha_0X_{ij}+U_{ij}$ within each group. This selection rule is used only to generate the data. Inference uses only the full equilibrium correspondence and remains agnostic about the equilibrium selection rule.

Let $Z_{B,i}:=\sum_{j\in G_B}X_{ij}$, $Z_{B,-ij}:=Z_{B,i}-X_{ij}$, $Z_{S,i}:=\sum_{j\in G_S}X_{ij}$, and $N_{S,i}:=\sum_{j\in G_S}Y_{ij}$. Our first auxiliary model, denoted as Model~1, combines a pooled regression for big-firm entry with a regression for the small-firm entry count:
\begin{align*}
Y_{ij}&=b_{B0}+b_{B1}X_{ij}+e_{ij},\qquad j\in G_B,\\
N_{S,i}&=b_{S0}+b_{S1}Z_{S,i}+\eta_i.
\end{align*}
Model 1 has four auxiliary coefficients $(b_{B0},b_{B1},b_{S0},b_{S1})$. It captures reduced-form relationships between big-firm entry and firms’ own characteristics, and between small-firm entry counts and aggregate small-firm characteristics. To capture additional variation associated with competition among firms, Model 2 augments these regressions with covariates describing rival firms:
\begin{align*}
Y_{ij}&=b_{B0}+b_{B1}X_{ij}+b_{B2}Z_{B,-ij}+b_{B3}Z_{S,i}+e_{ij},
\qquad j\in G_B,\\
N_{S,i}&=b_{S0}+b_{S1}Z_{S,i}+b_{S2}Z_{B,i}+\eta_i.
\end{align*}
Model 2 has seven auxiliary coefficients. As the results below show, the additional rival covariates in Model 2 yield smaller confidence intervals for the competition parameters. Both auxiliary models are estimated by ordinary least squares.

We use the weighted procedure described in Appendix~\ref{sec:weights}, estimating the full conditional score covariance separately in each sample and auxiliary model from five-fold cross-fitted conditional-mean predictions. Numerical delta critical values use $B=500$ Gaussian directions and $\epsilon_n=2n^{-1/4}$. Appendix~\ref{sec:simulation-details} gives the covariance construction and numerical details.

\subsection{Results}
Table~\ref{tab:simulation-coverage} reports pointwise coverage of the true parameter $\theta_0$ at the nominal $95\%$ level, using $5{,}000$ data repetitions for each $J\in \{10, 20, 50, 100\}$. Coverage ranges from \(96.4\%\) to \(98.2\%\), exceeding the nominal level in every design and indicating moderate conservativeness of our method in these experiments. Model 2 is more conservative than Model 1. Increasing the number of potential entrants from $10$ to $100$ does not produce undercoverage in these experiments.

\begin{table}[t]
\centering
\caption{Pointwise coverage and computation time}
\label{tab:simulation-coverage}
\small
\begin{tabular}{llcc}
\hline
$J$ & Auxiliary model & Coverage  & Inference time (seconds) \\
\hline
10 & Model 1 & 0.964  & 0.656 \\
10 & Model 2 & 0.966  & 2.952 \\[5pt]
20 & Model 1 & 0.965  & 1.163 \\
20 & Model 2 & 0.979  & 2.513 \\[5pt]
50 & Model 1 & 0.969  & 0.985 \\
50 & Model 2 & 0.982  & 1.650 \\[5pt]
100 & Model 1 & 0.969  & 1.100 \\
100 & Model 2 & 0.981  & 1.897 \\
\hline
\end{tabular}
\par\smallskip
\begin{minipage}{0.96\textwidth}
\footnotesize \textit{Notes:} Coverage is the fraction of $5{,}000$ repetitions in which the test does not reject $\theta_0=(1,1,4)$. Time averages $20$ sequential evaluations of one candidate parameter, including structural-support construction and the numerical delta test, but excluding data generation and estimation of the auxiliary coefficients and covariance.
\end{minipage}
\end{table}

The last column of Table~\ref{tab:simulation-coverage} reports the time required to test one candidate parameter on a Mac Studio equipped with an Apple M4 Max processor. These timings average 20 sequential, single-threaded evaluations at $\theta_0$. When constructing a confidence region, different candidate parameter values can be easily evaluated in parallel to further reduce the computation time. The results indicate that inference remains computationally practical for this game even with \(100\) potential entrants.

Table~\ref{tab:simulation-projections} reports the projected confidence intervals for the two auxiliary models under various values of $J$. These results are coordinate projections from a $51^3=132{,}651$-point Cartesian grid in the parameter space. For each sample, we take the smallest interval containing the nonrejected grid values of each coordinate. The table averages results across 100 random samples.

\begin{table}[t]
\centering
\caption{Mean projected interval endpoints and lengths}
\label{tab:simulation-projections}
\small
\setlength{\tabcolsep}{4pt}
\begin{tabular}{llcccccc}
\hline
 & & \multicolumn{2}{c}{$\alpha$} & \multicolumn{2}{c}{$\gamma_B$} & \multicolumn{2}{c}{$\gamma_S$} \\
$J$ & Model & Interval & Length & Interval & Length & Interval & Length \\
\hline
10 & 1 & $[0.934,1.726]$ & 0.792 & $[0.970,1.502]$ & 0.532 & $[0.607,13.017]$ & 12.410 \\
10 & 2 & $[0.973,1.554]$ & 0.581 & $[0.979,1.501]$ & 0.522 & $[3.490,6.003]$ & 2.513 \\[5pt]
20 & 1 & $[0.978,1.518]$ & 0.539 & $[0.911,1.374]$ & 0.463 & $[0.993,12.597]$ & 11.603 \\
20 & 2 & $[0.983,1.451]$ & 0.468 & $[0.981,1.351]$ & 0.369 & $[3.500,6.130]$ & 2.630 \\[5pt]
50 & 1 & $[0.990,1.447]$ & 0.457 & $[0.813,1.371]$ & 0.558 & $[0.723,13.443]$ & 12.720 \\
50 & 2 & $[0.988,1.411]$ & 0.423 & $[0.987,1.302]$ & 0.316 & $[3.747,5.813]$ & 2.067 \\[5pt]
100 & 1 & $[0.982,1.418]$ & 0.436 & $[0.758,1.359]$ & 0.601 & $[0.520,14.170]$ & 13.650 \\
100 & 2 & $[0.982,1.383]$ & 0.401 & $[0.985,1.270]$ & 0.286 & $[3.923,5.520]$ & 1.597 \\
\hline
\end{tabular}
\par\smallskip
\begin{minipage}{0.98\textwidth}
\footnotesize \textit{Notes:} Entries average across $100$ samples. ``Interval'' reports the mean lower and upper endpoints. Projections are interval hulls of nonrejected candidates on the grid specified in Appendix~\ref{sec:simulation-details}.
\end{minipage}
\end{table}

Table~\ref{tab:simulation-projections} shows that Model 2 produces shorter mean projected intervals than Model 1 for every parameter and market size. The largest gain concerns the competitive effect. At $J=100$, the mean length of the projected confidence interval for $\gamma_B$ falls from $0.601$ to $0.286$ and, for $\gamma_S$, it falls from $13.650$ to $1.597$. The reduction for \(\alpha\) is more modest, from $0.436$ to $0.401$. These results align with the intuition that additional rival covariates in Model 2 capture variation that is particularly informative about the competition parameters, yielding more informative inference on \(\gamma_B\) and \(\gamma_S\).

\section{Discussion}
\label{sec:discussion}

We conclude with a discussion of the interpretation of the procedure and of several extensions left for future work.

For complicated incomplete structural models, computing the sharp identified region is often computationally infeasible. A common response is to select a tractable subset of Artstein inequalities that is not core determining. The resulting outer region is generally nonsharp, and the identifying information lost by omitting inequalities can be difficult to interpret. Our indirect inference procedure also trades sharpness for tractability. The auxiliary estimating equations make the retained information explicit, by specifying which features of the observed data distribution a model compatible distribution must reproduce. As the three examples and the simulations illustrate, researchers can interpret these features through familiar outcome statistics and auxiliary regressions, and describe the additional data variation that the auxiliary comparison does not explicitly capture. Constructing an auxiliary model can therefore offer a more natural way to target informative data variation than selecting individual Artstein inequalities.

Enriching the auxiliary model by adding estimating equations and retaining the existing restrictions weakly shrinks the outer identified region $\Theta_n(s)$ and can reduce the loss of sharpness. This suggests an extension in which the dimension $d_\beta$ of the auxiliary parameter grows with the sample size, allowing the auxiliary model to capture progressively more features of the data distribution. Asymptotic theory for $M$-estimators with a growing number of parameters, as in \cite{Portnoy:85}, provides a natural starting point. Establishing valid inference in this setting, together with conditions under which the resulting outer regions approach the sharp identified region, is left for future work.

Applied work with structural general equilibrium models frequently combines targeted moments with auxiliary regressions, as in \cite{Tintelnot:2017} and \cite{Gaubert:2018}. Since our framework accommodates arbitrary estimating equations $s(y\vert x;\beta)$, including stacked moments and regression scores, it may provide a formal justification for a variant of this practice in settings where the underlying structural model is incomplete.

\begin{appendix}

\section{Cross-fit strategy for Assumption~\ref{ass:boot}}
\label{sec:cross-fit}

We propose a construction of a sequence~$\mathbb Z_n^\ast$ that satisfies Assumption~\ref{ass:boot} in the just-identified case $d_s=d_\beta$ with~$r_n=\sqrt{n}$. Throughout this appendix, the population and sample Jacobians used below are assumed nonsingular whenever their inverses are taken. The inverse-Jacobian construction is not asserted for overidentified estimating equations, which require an appropriate GMM influence matrix. Write
\begin{eqnarray*}
\mu_s(x;\beta) \; := \; \mathbb E_{P_{Y\vert X,0}}\left[ s(Y\vert x;\beta)\;\vert\;X=x\right],
& \mbox{ and } &
\xi_{n,i} \; := \; s(Y_i\vert X_i;\beta_n)-\mu_s(X_i;\beta_n).
\end{eqnarray*}
By the definition of the pseudo-true value in equation~(\ref{eq:aux}), we have~$n^{-1}\sum_{i=1}^n\mu_s(X_i;\beta_n)=0$, and therefore
\begin{eqnarray}
\label{eq:exact-residual}
\sum_{i=1}^n s(Y_i\vert X_i;\beta_n) & = & \sum_{i=1}^n\xi_{n,i}.
\end{eqnarray}
Expanding the sample estimating equations around~$\beta_n$ and using~(\ref{eq:exact-residual}) gives
\begin{eqnarray}
\label{eq:influence}
\sqrt{n}\left(\hat\beta_n-\beta_n\right) & = &
H_n\;\frac{1}{\sqrt{n}}\sum_{i=1}^n\xi_{n,i}+o_p(1),
\end{eqnarray}
where~$A_n:=n^{-1}\sum_{i=1}^n\nabla_\beta\mu_s(X_i;\beta_n)$ and~$H_n:=-A_n^{-1}$. The summands
are conditionally independent and conditionally centered, so the conditional
variance of the right hand side is~$H_nB_{c,n}H_n^\top$ with
\begin{eqnarray*}
B_{c,n} & := & \frac1n\sum_{i=1}^n\mathbb E\left[
\xi_{n,i}\xi_{n,i}^\top\;\vert\;X_i\right].
\end{eqnarray*}
Hence, Assumption~\ref{ass:est} holds by the conditional Lindeberg theorem
with~$\mathbb Z_0$ centered Gaussian of variance~$\Omega_c:=HB_cH^\top$,
where~$H$ and~$B_c$ are the almost sure limits of~$H_n$ and~$B_{c,n}$.

We construct~$\mathbb Z_n^\ast$ from residualized estimating equations, with the conditional mean~$\mu_s$ estimated nonparametrically and cross-fitted. Partition~$\{1,\ldots,n\}$ into~$K\geq2$ folds~$I_1,\ldots,I_K$ of comparable size, write~$k(i)$ for the fold containing observation~$i$, and let~$\hat\beta_n^{(-k)}$ and~$\hat\mu_s^{(-k)}$ denote, respectively, the auxiliary estimator and a nonparametric estimator of~$\mu_s$, both computed from the observations outside fold~$k$ only. Set
\begin{eqnarray}
\label{eq:xihat}
\hat\xi_i & := & s\left( Y_i\vert X_i;\hat\beta_n^{(-k(i))}\right)-\hat\mu_s^{(-k(i))}\left( X_i;\hat\beta_n^{(-k(i))}\right), \qquad i=1,\ldots,n,
\end{eqnarray}
and
\begin{eqnarray*}
    \hat\Omega_c:=\hat H_n\hat B_{c,n}\hat H_n^\top, & \mbox{with} & \hat B_{c,n}:=\frac1n\sum_{i=1}^n\hat\xi_i\hat\xi_i^\top,\;\;
    \hat H_n \; := \; -\hat A_n^{-1},\;\;
\hat A_n:=\frac1n\sum_{i=1}^n\nabla_\beta s\left( Y_i\vert X_i;\hat\beta_n\right).
\end{eqnarray*}
Finally, draw~$\mathbb Z_n^\ast$ directly from the~$N(0,\hat\Omega_c)$ distribution. The following lemma gives primitive conditions under which this construction satisfies Assumption~\ref{ass:boot}, and under which $B_{c,n}$ and $\hat B_{c,n}$ deliver weights satisfying Assumption~\ref{ass:weights} of Appendix~\ref{sec:weights}. Conditions~(a) and~(b) are those under which Assumption~\ref{ass:est} was obtained above, and are statements about almost every covariate sequence; condition~(c), which concerns the cross-fitted nuisance estimates, is only required unconditionally, in probability with respect to the joint distribution of the data.

\begin{lemma}
\label{lem:crossfit}
Suppose that, almost surely with respect to the distribution of $(X_i)_{i=1}^\infty$: (a) $B_{c,n}\rightarrow B_c$ with $B_c$ positive definite, and $A_n\rightarrow A$ with $A$ nonsingular; and (b) for every $\eta>0$, $n^{-1}\sum_{i=1}^n\mathbb E[\|\xi_{n,i}\|^2\mathbf 1\{\|\xi_{n,i}\|>\eta\sqrt n\}\,\vert\,X_i]\rightarrow0$. Suppose also that (c) $n^{-1}\sum_{i=1}^n\|\hat\xi_i-\xi_{n,i}\|^2=o_p(1)$ and $\hat A_n-A_n=o_p(1)$. Then $\hat B_{c,n}-B_{c,n}=o_p(1)$, $\hat\Omega_c-\Omega_c=o_p(1)$, and $\mathbb Z_n^\ast\sim N(0,\hat\Omega_c)$ satisfies Assumption~\ref{ass:boot}. Moreover, let $g$ map the symmetric $d_s\times d_s$ matrices to the nonsingular ones, continuously on the positive definite matrices, for instance $g(B)=B^{-1/2}$ or the inverse of the Cholesky factor of $B$ when $B$ is positive definite, and $g(B)=I_{d_s}$ otherwise. Then Assumption~\ref{ass:weights} holds with $W_n:=g(B_{c,n})$ and $\hat W_n:=g(\hat B_{c,n})$.
\end{lemma}

Since the weighted binding function of Appendix~\ref{sec:weights} depends on $W$ only through $W^\top W$, all choices of $g$ with $g(B)^\top g(B)=B^{-1}$ on the positive definite matrices, such as the two mentioned in the lemma, yield the same weighted binding function; the Cholesky choice is the one implemented in Appendix~\ref{sec:simulation-details}.

\begin{proof}[Proof of Lemma~\ref{lem:crossfit}]
\emph{Step 1.} Fix a covariate sequence in the almost sure set of conditions~(a) and~(b). In this step, probabilities and expectations are conditional on it, so that the $\xi_{n,i}$ are independent across $i$ with mean zero, and we write $\Sigma_{n,i}:=\mathbb E[\xi_{n,i}\xi_{n,i}^\top\vert X_i]$, so that $n^{-1}\sum_{i=1}^n\Sigma_{n,i}=B_{c,n}$. Fix $\eta>0$ and write $\xi_{n,i}=\xi^\leq_{n,i}+\xi^>_{n,i}$, with $\xi^\leq_{n,i}:=\xi_{n,i}\mathbf 1\{\|\xi_{n,i}\|\leq\eta\sqrt n\}$. Since the two indicators are disjoint, $n^{-1}\sum_{i=1}^n(\xi_{n,i}\xi_{n,i}^\top-\Sigma_{n,i})=M_n^\leq+M_n^>$, where $M_n^\leq:=n^{-1}\sum_{i=1}^n(\xi^\leq_{n,i}\xi^{\leq\top}_{n,i}-\mathbb E[\xi^\leq_{n,i}\xi^{\leq\top}_{n,i}\vert X_i])$ and $M_n^>$ is defined in the same way from $\xi^>_{n,i}$. The summands of $M_n^\leq$ are independent and centered, and $\sum_{k,l}(\xi_k\xi_l)^2=\|\xi\|^4$ for any vector $\xi$, so that, with $\|\cdot\|_F$ the Frobenius norm,
\begin{eqnarray*}
\mathbb E\left[\|M_n^\leq\|_F^2\right] \;\leq\; \frac{1}{n^2}\sum_{i=1}^n\mathbb E\left[\|\xi^\leq_{n,i}\|^4\right] \;\leq\; \frac{1}{n^2}\sum_{i=1}^n\eta^2n\,\mathbb E\left[\|\xi_{n,i}\|^2\right] \;=\; \eta^2\,\mathrm{tr}B_{c,n},
\end{eqnarray*}
and Chebyshev's inequality gives $\mathbb P(\|M_n^\leq\|_F>\epsilon)\leq\eta^2\mathrm{tr}B_{c,n}/\epsilon^2$. By condition~(b), $\mathbb E\|M_n^>\|\leq2n^{-1}\sum_{i=1}^n\mathbb E[\|\xi_{n,i}\|^2\mathbf 1\{\|\xi_{n,i}\|>\eta\sqrt n\}]\rightarrow0$, so that $M_n^>\rightarrow0$ in probability by Markov's inequality. Hence, by condition~(a), $\limsup_n\mathbb P(\|n^{-1}\sum_{i=1}^n(\xi_{n,i}\xi_{n,i}^\top-\Sigma_{n,i})\|>2\epsilon)\leq\eta^2\mathrm{tr}B_c/\epsilon^2$ for every $\eta>0$, and letting $\eta\downarrow0$ shows that $n^{-1}\sum_{i=1}^n\xi_{n,i}\xi_{n,i}^\top-B_{c,n}\rightarrow0$ in probability conditionally on the covariate sequence, almost surely, hence also unconditionally, by dominated convergence.

\emph{Step 2.} Since $\hat\xi\hat\xi^\top-\xi\xi^\top=(\hat\xi-\xi)(\hat\xi-\xi)^\top+(\hat\xi-\xi)\xi^\top+\xi(\hat\xi-\xi)^\top$, the Cauchy--Schwarz inequality gives
\begin{eqnarray*}
\left\|\hat B_{c,n}-\frac1n\sum_{i=1}^n\xi_{n,i}\xi_{n,i}^\top\right\| & \leq & 2\left(\frac1n\sum_{i=1}^n\|\xi_{n,i}\|^2\right)^{1/2}\left(\frac1n\sum_{i=1}^n\|\hat\xi_i-\xi_{n,i}\|^2\right)^{1/2}+\frac1n\sum_{i=1}^n\|\hat\xi_i-\xi_{n,i}\|^2.
\end{eqnarray*}
The conditional expectation of $n^{-1}\sum_{i=1}^n\|\xi_{n,i}\|^2$ given the covariates is $\mathrm{tr}B_{c,n}$, which converges almost surely by condition~(a), so that $n^{-1}\sum_{i=1}^n\|\xi_{n,i}\|^2=O_p(1)$ by the conditional Markov inequality and dominated convergence. Together with condition~(c) and Step~1, this gives $\hat B_{c,n}-B_{c,n}=o_p(1)$.

\emph{Step 3.} By conditions~(a) and~(c), $\hat A_n\rightarrow A$ in probability, so that $\hat H_n\rightarrow H$ by continuity of matrix inversion at the nonsingular $A$, and $\hat\Omega_c=\hat H_n\hat B_{c,n}\hat H_n^\top\rightarrow HB_cH^\top=\Omega_c$ in probability by Step~2 and condition~(a). Let $G\sim N(0,I_{d_\beta})$ be independent of the data. Conditionally on the data, $\mathbb Z_n^\ast$ has the law of $\hat\Omega_c^{1/2}G$, while $\mathbb Z_0$ has the law of $\Omega_c^{1/2}G$, so that, for every $f\in\mbox{BL}_1(\real^{d_\beta})$,
\begin{eqnarray*}
\left|\mathbb E\left[f(\mathbb Z_n^\ast)\,\vert\,(X_i,Y_i)_{i=1}^n\right]-\mathbb E\left[f(\mathbb Z_0)\,\vert\,(X_i)_{i=1}^n\right]\right|
& \leq & \mathbb E\left[\min\left\{2,\|(\hat\Omega_c^{1/2}-\Omega_c^{1/2})G\|\right\}\,\Big\vert\,(X_i,Y_i)_{i=1}^n\right] \\
& \leq & \|\hat\Omega_c^{1/2}-\Omega_c^{1/2}\|\,\mathbb E\|G\|,
\end{eqnarray*}
uniformly in $f$. The right-hand side is $o_p(1)$ by continuity of the square root on the positive semidefinite matrices, which is Assumption~\ref{ass:boot}.

\emph{Step 4.} The matrices $W_n=g(B_{c,n})$ are nonsingular and measurable functions of $(X_i)_{i=1}^n$. Almost surely, $B_{c,n}\rightarrow B_c$ with $B_c$ positive definite by condition~(a), so that $(B_{c,n})_{n\geq n_1}$ lies in a compact subset of the positive definite matrices for some finite $n_1$, on which $g$ is bounded by continuity; hence $\sup_n\|W_n\|<\infty$ almost surely. By Step~2 and condition~(a), $\hat B_{c,n}\rightarrow B_c$ in probability, so that $\hat B_{c,n}$ is positive definite with probability tending to one and, by continuity of $g$ at $B_c$ and of matrix inversion at $g(B_c)$, $\hat W_nW_n^{-1}=g(\hat B_{c,n})g(B_{c,n})^{-1}\rightarrow g(B_c)g(B_c)^{-1}=I_{d_s}$ in probability, which is Assumption~\ref{ass:weights}.
\end{proof}

\section{Weighted estimating equations and estimated weights}
\label{sec:weights}

The binding function of Definition~\ref{def:bind-n} depends on the scale and orientation of the auxiliary estimating equations. The finite sample performance of our procedure may be poor if the moment functions in the auxiliary model vary wildly in scale. We therefore propose a variant of our procedure that allows reweighting of the auxiliary moment functions with an estimated variance-covariance matrix. To that end, we define the weighted binding function
\begin{eqnarray}
\label{eq:bindW}
L_n(\theta,\beta;W) & := & \inf \left\{ \left\| W\,
\frac1n\sum_{i=1}^n\mathbb E_{P_{Y\vert X}} \left[ s(Y\vert X_i;\beta)\; \vert\; X_i\right] \right\|
: \; P_{Y\vert X} \in \mathcal P_\theta \right\}.
\end{eqnarray}
By construction, $L_n(\theta,\beta;I_{d_s})$ is the binding function~$L_n(\theta,\beta)$ of Definition~\ref{def:bind-n}, and $L_n(\theta,\beta;W)$ is the unweighted binding function associated with the transformed estimating function $(y,x,\beta)\mapsto Ws(y\vert x;\beta)$, which satisfies Assumption~\ref{ass:aux} with the same pseudo-true value $\beta_n$. The weighting matrix plays the role of the weighting matrix in traditional indirect inference \citep{GMR:93,GT:96}. We recommend~$W_n:=B_{c,n}^{-1/2}$, where $B_{c,n}$ is the conditional variance of the centered estimating equations introduced in Appendix~\ref{sec:cross-fit}. That way, the weighted estimating equations are standardized given the covariates and the weighted binding function is invariant to nonsingular linear transformations of $s$. Such weights are unknown and must be estimated. We show in this appendix that the results of Sections~\ref{sec:ii} and~\ref{sec:inference} hold for weighted binding functions, and that they are unaffected by the replacement of the weights with consistent estimates.

For a nonsingular $d_s\times d_s$ matrix $W$, $\theta\in\Theta$, $\beta\in\mathbb B_0$ and $\lambda\in\mathcal S$, write $Q_i:=Q_{U\vert X_i;\theta_2}$ and
\begin{eqnarray}
\label{eq:phinW}
\varphi_n(\lambda,\beta;W) & := & \frac1n\sum_{i=1}^n \mathbb E_{Q_i}\left[ \inf_{y\in\Gamma(U\vert X_i;\theta_1)} \lambda^\top Ws(y\vert X_i;\beta)\;\Big\vert\; X_i\right] \; = \; \varphi_n(W^\top\lambda,\beta),
\end{eqnarray}
where $\varphi_n$ is the dual objective of~(\ref{eq:phin}). $\hat\Lambda_n(\theta,\beta;W):=\argmax_{\lambda\in\mathcal S}\varphi_n(\lambda,\beta;W)$ is the weighted outer solution set, and $\tilde L_n(\theta,\beta;W):=L_n(\theta,\Pi_{\mathbb B_0}(\beta);W)$, $\beta\in\real^{d_\beta}$, is the projection extension of Definition~\ref{def:ext}. The weighted inner program in direction $\lambda$ is the unweighted inner program in direction $W^\top\lambda$, since $\lambda^\top Ws=(W^\top\lambda)^\top s$, so its solution set is $\hat{\mathcal Y}_i(u,W^\top\lambda,\beta)$ in the notation of Section~\ref{sec:ii}. Finally, for nonsingular $W$ and $W^\prime$, let
\begin{eqnarray*}
\kappa(W,W^\prime) & := & \max\left\{ \left\|W^\prime W^{-1}\right\|,\;\left\|WW^{\prime-1}\right\| \right\} \; \geq \; 1,
\end{eqnarray*}
the inequality holding because $\|W^\prime W^{-1}\|\,\|WW^{\prime-1}\|\geq\|I_{d_s}\|=1$. We say that Assumption~\ref{ass:uniq} holds for the weight $W$ if it holds with the estimating function $s$ replaced by $s^W:=Ws$. Likewise, for a sequence $(W_n)_{n\geq1}$ of nonsingular matrices, Assumptions~\ref{ass:unif} and~\ref{ass:cv} hold for the weight sequence $(W_n)$ if they hold with $L_n(\theta,\cdot)$ and $L^\prime_{\beta_n}(\theta,\cdot)$ replaced by $L_n(\theta,\cdot;W_n)$ and by the weighted directional derivative $L^\prime_{\beta_n}(\theta,\cdot;W_n)$ of~(\ref{eq:hddW}) below.

\begin{proposition}
\label{prop:weighted}
Let Assumption~\ref{ass:reg} hold, and let $W$ and $W^\prime$ be nonsingular $d_s\times d_s$ matrices. Then, for each $n\geq1$ and all realizations $(X_1,\ldots,X_n)$:
\begin{enumerate}[label=(\roman*)]
\item For all $(\theta,\beta)\in\Theta\times\mathbb B_0$,
\begin{eqnarray}
\label{eq:comparison}
\kappa(W,W^\prime)^{-1}\,L_n(\theta,\beta;W) \;\leq\; L_n(\theta,\beta;W^\prime) \;\leq\; \kappa(W,W^\prime)\,L_n(\theta,\beta;W),
\end{eqnarray}
and the same inequalities hold between $\tilde L_n(\theta,\beta;W)$ and $\tilde L_n(\theta,\beta;W^\prime)$ for all $\beta\in\real^{d_\beta}$. In particular, $L_n(\theta,\beta;W)=0$ if and only if $L_n(\theta,\beta)=0$, so that, under Assumption~\ref{ass:aux}, $L_n(\theta,\beta_n;W)=0$ for each $\theta\in\Theta_I$, almost surely with respect to the distribution of $(X_i)_{i=1}^\infty$.
\item $\forall(\theta,\beta)\in\Theta\times\mathbb B_0$, $L_n(\theta,\beta;W)=\sup_{\lambda\in\mathcal S}\varphi_n(\lambda,\beta;W)$, and $\hat\Lambda_n(\theta,\beta;W)$ is nonempty and compact.
\end{enumerate}
If, in addition, Assumption~\ref{ass:aux} holds, then, for each $\theta\in\Theta$, almost surely with respect to the distribution of $(X_i)_{i=1}^\infty$:
\begin{enumerate}[label=(\roman*),resume]
\item $\beta\mapsto L_n(\theta,\beta;W)$ is Lipschitz on $\mathbb B_0$, and $\beta\mapsto\tilde L_n(\theta,\beta;W)$ is Lipschitz on $\real^{d_\beta}$, with constant $\|W\|\ell_n$, where $\ell_n$ is the constant of Proposition~\ref{prop:hdd}(i). If Assumption~\ref{ass:uniq} holds for the weight $W$, then, for all sufficiently large $n$, $\beta\mapsto L_n(\theta,\beta;W)$ and $\beta\mapsto\tilde L_n(\theta,\beta;W)$ are Hadamard directionally differentiable at $\beta_n$, with derivative
\begin{eqnarray}
\label{eq:hddW}
L^\prime_{\beta_n}(\theta,d;W) & = & \max_{\lambda\in\hat\Lambda_n(\theta,\beta_n;W)}\;\frac1n\sum_{i=1}^n \mathbb E_{Q_i}\left[ \min_{y\in\hat{\mathcal Y}_i(U,W^\top\lambda,\beta_n)} \lambda^\top W\nabla_\beta s(y\vert X_i;\beta_n)\, d \;\Big\vert\; X_i\right]
\end{eqnarray}
for every direction $d\in\real^{d_\beta}$.
\end{enumerate}
\end{proposition}

Parts~(i)--(iii) say that Propositions~\ref{prop:idset}--\ref{prop:hdd} hold for the weighted binding function, with the inner programs solved in the transformed directions $W^\top\lambda$ and the Lipschitz constant inflated by $\|W\|$. The comparison inequality~(\ref{eq:comparison}) is the key to the treatment of estimated weights: it is multiplicative, not additive, so that a weight $W^\prime$ close to $W$ in the relative sense $\kappa(W,W^\prime)\approx1$ perturbs the binding function by a factor close to one uniformly in $(\theta,\beta)$, and in particular leaves it close to zero wherever it is close to zero. 

\subsection*{Estimated weights}

We now replace the weights with estimates.

\begin{assumption}[Weights]
\label{ass:weights}
$(W_n)_{n\geq1}$ is a sequence of nonsingular $d_s\times d_s$ matrices, with $W_n$ a measurable function of $(X_i)_{i=1}^n$, such that $\sup_{n\geq1}\|W_n\|<\infty$ almost surely with respect to the distribution of $(X_i)_{i=1}^\infty$; and $(\hat W_n)_{n\geq1}$ is a sequence of $d_s\times d_s$ matrices, with $\hat W_n$ a measurable function of $(Y_i,X_i)_{i=1}^n$, such that~$\|\hat W_nW_n^{-1}-I_{d_s} \| \rightarrow 0$ in probability.
\end{assumption}

The population weights $W_n$ may depend on the covariate design, as the conditional variance $B_{c,n}$ does, and are treated as fixed conditionally on the covariates, in the same way as the binding function $L_n$ and the pseudo-true value $\beta_n$. As in Assumption~\ref{ass:boot}, convergence in probability of $\hat W_nW_n^{-1}$ is with respect to the joint distribution of the data. Since the weighted binding function~(\ref{eq:bindW}) depends on $W$ only through $W^\top W$, any choice of $W_n$ with $W_n^\top W_n=B_{c,n}^{-1}$, such as $B_{c,n}^{-1/2}$ or the inverse of the Cholesky factor of $B_{c,n}$ used in Appendix~\ref{sec:simulation-details}, yields the same binding function, and Lemma~\ref{lem:crossfit} in Appendix~\ref{sec:cross-fit} shows that such weights and their cross-fitted estimates $\hat W_n$ satisfy Assumption~\ref{ass:weights} whenever the limit $B_c$ of $B_{c,n}$ is positive definite.

The feasible test statistic and numerical directional derivative replace $W_n$ with $\hat W_n$ in the binding function: for $\theta\in\Theta$ and $h\in\real^{d_\beta}$, with $\epsilon_n$ as in Definition~\ref{def:ndd},
\begin{eqnarray}
\label{eq:feasible}
\hat T_n(\theta) & := & r_nL_n(\theta,\hat\beta_n;\hat W_n), \nonumber \\
\hat L^\prime_{\beta_n}(\theta,h;\hat W_n) & := & \frac{1}{\epsilon_n}\left( L_n(\theta,\hat\beta_n+\epsilon_nh;\hat W_n)-L_n(\theta,\hat\beta_n;\hat W_n)\right),
\end{eqnarray}
and, for $\tau\in(0,1)$, the $\tau$-quantile of the conditional distribution, given the data, of the random variable $\hat L^\prime_{\beta_n}(\theta,\mathbb Z_n^\ast;\hat W_n)$ is denoted $\hat c_{n,\tau}(\theta;\hat W_n)$. The resampling draws $\mathbb Z_n^\ast$ may be generated from the same estimated quantities as $\hat W_n$, for instance from $N(0,\hat\Omega_c)$ with $\hat\Omega_c$ computed from $\hat B_{c,n}$ as in Appendix~\ref{sec:cross-fit}: no independence between $\hat W_n$ and $\mathbb Z_n^\ast$ is required, only Assumptions~\ref{ass:boot} and~\ref{ass:weights}.

\begin{theorem}
\label{thm:weights}
Fix $\alpha\in(0,1)$ for which Assumption~\ref{ass:cv} holds for the weight sequence $(W_n)$. Let Assumptions~\ref{ass:aux}--\ref{ass:reg} and~\ref{ass:weights} hold, let Assumption~\ref{ass:uniq} hold for the weight $W_n$ for each $n\geq1$, and let Assumption~\ref{ass:unif} hold for the weight sequence $(W_n)$. Then, for each $\theta\in\Theta_I$,
\begin{eqnarray*}
\limsup_n\;\mathbb P\left( \hat T_n(\theta)>\hat c_{n,1-\alpha}(\theta;\hat W_n)\right) & \leq & \alpha.
\end{eqnarray*}
\end{theorem}

The confidence region $\{\theta\in\Theta:\;\hat T_n(\theta)\leq\hat c_{n,1-\alpha}(\theta;\hat W_n)\}$ therefore covers each point of the sharp identified region with asymptotic probability at least $1-\alpha$, exactly as $\mathcal C_{n,1-\alpha}$ does. The proof, at the end of this appendix, shows that the estimation error in the weights perturbs the test statistic and the numerical directional derivatives by factors $\kappa(W_n,\hat W_n)\rightarrow1$. This multiplicative perturbation is negligible without rate conditions on~$\hat W_n$.

\subsection*{Proofs}

\begin{proof}[Proof of Proposition~\ref{prop:weighted}]
Fix $n$, $(X_1,\ldots,X_n)$ and $\theta\in\Theta$, and write $Q_i:=Q_{U\vert X_i;\theta_2}$. For $P_{Y\vert X}\in\mathcal P_\theta$ and $\beta\in\mathbb B_0$, let $m(P_{Y\vert X},\beta):=\frac1n\sum_{i=1}^n\mathbb E_{P_{Y\vert X}}[s(Y\vert X_i;\beta)\vert X_i]$, which is well defined by Assumption~\ref{ass:reg}(c), so that $L_n(\theta,\beta;W)=\inf\{\|Wm(P_{Y\vert X},\beta)\|:\;P_{Y\vert X}\in\mathcal P_\theta\}$, the infimum being over a nonempty set by Assumption~\ref{ass:reg}(a).

\medskip\noindent\emph{Part (i).} For every $\mu\in\real^{d_s}$, $\|W^\prime\mu\|=\|W^\prime W^{-1}W\mu\|\leq\|W^\prime W^{-1}\|\,\|W\mu\|\leq\kappa(W,W^\prime)\,\|W\mu\|$. Applied to $\mu=m(P_{Y\vert X},\beta)$, followed by the infimum over $\mathcal P_\theta$, this gives the right inequality in~(\ref{eq:comparison}); exchanging the roles of $W$ and $W^\prime$ gives the left one. The inequalities for the extensions follow by evaluating~(\ref{eq:comparison}) at $\Pi_{\mathbb B_0}(\beta)$. Taking $W^\prime=I_{d_s}$ shows that $L_n(\theta,\beta;W)$ and $L_n(\theta,\beta)$ vanish together, and Proposition~\ref{prop:idset} completes the proof of part~(i).

\medskip\noindent\emph{Parts (ii) and (iii).} Let $s^W(y\vert x;\beta):=Ws(y\vert x;\beta)$. Since $W$ is nonsingular, $s^W$ satisfies Assumption~\ref{ass:aux} with the same pseudo-true value $\beta_n$. It satisfies Assumption~\ref{ass:reg}(b) by linearity, with $\nabla_\beta s^W=W\nabla_\beta s$, and Assumption~\ref{ass:reg}(c) with envelope $\|W\|\,b(u,x;\theta)$, since $\|Ws\|\leq\|W\|\|s\|$ and $\|W\nabla_\beta s\|\leq\|W\|\|\nabla_\beta s\|$; parts~(a) and~(d) of Assumption~\ref{ass:reg} do not involve $s$. The unweighted binding function associated with $s^W$ is $L_n(\theta,\beta;W)$, and, since $\lambda^\top s^W=(W^\top\lambda)^\top s$, its dual objective is $\varphi_n(W^\top\lambda,\beta)=\varphi_n(\lambda,\beta;W)$, its inner solution set in direction $\lambda$ is $\hat{\mathcal Y}_i(u,W^\top\lambda,\beta)$, and its outer solution set is $\hat\Lambda_n(\theta,\beta;W)$. Part~(ii) is therefore Proposition~\ref{prop:dual} applied to $s^W$, and the nonemptiness and compactness of $\hat\Lambda_n(\theta,\beta;W)$ follow from the continuity of $\lambda\mapsto\varphi_n(\lambda,\beta;W)$ on the compact set $\mathcal S$, established in Step~1 of the proof of Proposition~\ref{prop:hdd} applied to $s^W$, which only uses the finiteness of $\frac1n\sum_{i=1}^n\mathbb E_{Q_i}[b(U,X_i;\theta)]$ under Assumption~\ref{ass:reg}(c). Part~(iii) is Proposition~\ref{prop:hdd} applied to $s^W$: the Lipschitz constant is $\frac1n\sum_{i=1}^n\mathbb E_{Q_i}[\|W\|b(U,X_i;\theta)]=\|W\|\ell_n$; the condition of Proposition~\ref{prop:hdd} is Assumption~\ref{ass:uniq} for the weight $W$; and the derivative formula~(\ref{eq:hdd}) for $s^W$ is~(\ref{eq:hddW}), the gradient of $s^W$ being $W\nabla_\beta s$. The statements for the extension follow as in the discussion of Definition~\ref{def:ext}: $\Pi_{\mathbb B_0}$ is $1$-Lipschitz, and $\tilde L_n(\theta,\cdot;W)$ coincides with $L_n(\theta,\cdot;W)$ on the ball of radius $\delta_\ast$ around $\beta_n$, which lies in $\mathbb B_0$ for $n$ large enough by Assumption~\ref{ass:reg}(d).

\end{proof}

\begin{proof}[Proof of Theorem~\ref{thm:weights}]
Fix $\theta\in\Theta_I$ and $\alpha\in(0,1)$. As in the proof of Theorem~\ref{thm:ndm}, we first reduce the claim to a conditional statement along a subsequence. By Lemma~\ref{lem:transfer}(i), it suffices to show that every subsequence of $(n)_{n\geq1}$ has a further subsequence along which $\limsup_n\mathbb P(\hat T_n(\theta)>\hat c_{n,1-\alpha}(\theta;\hat W_n)\,\vert\,(X_i)_{i=1}^\infty)\leq\alpha$ almost surely. By Lemma~\ref{lem:transfer}(ii) applied to the sum of the left-hand side $\rho_n^{\mathbb Z}$ of Assumption~\ref{ass:boot} and of $\|\hat W_nW_n^{-1}-I_{d_s}\|$, every subsequence has a further subsequence along which both quantities converge to zero in probability conditionally on $(X_i)_{i=1}^\infty$, almost surely with respect to the distribution of $(X_i)_{i=1}^\infty$; the remaining assumptions are limit statements along the full sequence and hold along every subsequence. We fix such a subsequence, along which all limits below are taken and whose elements we still denote by $n$, and condition throughout on a realization of $(X_i)_{i=1}^\infty$ in the almost sure set on which Assumptions~\ref{ass:aux}--\ref{ass:reg}, Assumption~\ref{ass:uniq} for the weights $W_n$, Assumptions~\ref{ass:unif} and~\ref{ass:cv} for the weight sequence $(W_n)$, and the conclusions of Proposition~\ref{prop:weighted} hold, on which $\rho_n^{\mathbb Z}\rightarrow0$ and $\|\hat W_nW_n^{-1}-I_{d_s}\|\rightarrow0$ in probability conditionally on the covariates along the subsequence, and on which $\bar\ell_W:=\sup_n\|W_n\|\ell_n<\infty$, which is the case almost surely since $\sup_n\|W_n\|<\infty$ and $\sup_n\ell_n<\infty$ almost surely. All binding functions are understood as their extensions, and all probability statements are conditional on the covariates, with randomness coming from $(Y_i)_{i=1}^n$ and the resampling scheme. Write $Z_n:=r_n(\hat\beta_n-\beta_n)$; $T_n:=r_nL_n(\theta,\hat\beta_n;W_n)$ for the infeasible test statistic with population weights; $L^\prime_n(h):=L^\prime_{\beta_n}(\theta,h;W_n)$ for the weighted directional derivative~(\ref{eq:hddW}); $\hat L^\prime_n(h):=\epsilon_n^{-1}(L_n(\theta,\hat\beta_n+\epsilon_nh;W_n)-L_n(\theta,\hat\beta_n;W_n))$ and $\hat L^\prime_n(h;\hat W_n):=\hat L^\prime_{\beta_n}(\theta,h;\hat W_n)$ of~(\ref{eq:feasible}) for the numerical derivatives with population and estimated weights; $c_n$ for the $(1-\alpha)$-quantile of $L^\prime_n(\mathbb Z_0)$; and $\hat c_n:=\hat c_{n,1-\alpha}(\theta;\hat W_n)$.

By Proposition~\ref{prop:weighted}, the binding function $L_n(\theta,\cdot;W_n)$ satisfies the conclusions of Propositions~\ref{prop:idset}, \ref{prop:dual} and~\ref{prop:hdd}, with Lipschitz constants $\|W_n\|\ell_n\leq\bar\ell_W$, and Assumptions~\ref{ass:unif} and~\ref{ass:cv} hold for it by hypothesis. These, together with Assumption~\ref{ass:est} and the convergence $\rho_n^{\mathbb Z}\rightarrow0$ in probability conditionally on the covariates, are the only properties used in the proof of Theorem~\ref{thm:ndm} once the covariates are fixed, which therefore applies verbatim to $T_n$, $\hat L^\prime_n$ and $c_n$ along the subsequence; in particular, $(Z_n)$ is uniformly tight, $(\mathbb Z_n^\ast)$ is uniformly tight in probability, and~(\ref{eq:rho-n}) and~(\ref{eq:rho-star}) hold for $T_n$ and $\hat L^\prime_n(\mathbb Z_n^\ast)$, with $L^\prime_n$ the weighted derivative. It remains to show that the replacement of $W_n$ with $\hat W_n$ perturbs the test statistic and the numerical derivatives by asymptotically negligible amounts. We proceed in three steps.

\medskip\noindent\emph{Step 1: } Let $\Delta_n:=\hat W_nW_n^{-1}-I_{d_s}$, so that $\|\Delta_n\|\rightarrow0$ in probability by the choice of the subsequence, and let $G_n:=\{\|\Delta_n\|\leq1/2\}$, so that $\mathbb P(G_n)\rightarrow1$. On $G_n$, the matrix $\hat W_nW_n^{-1}=I_{d_s}+\Delta_n$ is nonsingular, hence so is $\hat W_n$, with $\|\hat W_nW_n^{-1}\|\leq1+\|\Delta_n\|$ and, by the Neumann series, $\|W_n\hat W_n^{-1}\|=\|(I_{d_s}+\Delta_n)^{-1}\|\leq(1-\|\Delta_n\|)^{-1}\leq1+2\|\Delta_n\|$. Hence, on $G_n$,
\begin{eqnarray}
\label{eq:kappa-n}
1 \;\leq\; \kappa_n \;:=\; \kappa(W_n,\hat W_n) \;\leq\; 1+2\|\Delta_n\|,
\end{eqnarray}
so that $(\kappa_n-1)\mathbf 1_{G_n}\rightarrow0$ in probability. Moreover, on $G_n$, Proposition~\ref{prop:weighted}(i) applied to the extensions gives, for every $\beta\in\real^{d_\beta}$,
\begin{eqnarray}
\label{eq:relative}
\left| L_n(\theta,\beta;\hat W_n)-L_n(\theta,\beta;W_n)\right| & \leq & (\kappa_n-1)\,L_n(\theta,\beta;W_n) \nonumber \\
& \leq & (\kappa_n-1)\,\bar\ell_W\,\|\beta-\beta_n\|,
\end{eqnarray}
where the first inequality follows from~(\ref{eq:comparison}), which bounds the difference from above by $(\kappa_n-1)L_n(\theta,\beta;W_n)$ and from below by $-(1-\kappa_n^{-1})L_n(\theta,\beta;W_n)$, together with the elementary inequality $1-\kappa_n^{-1}\leq\kappa_n-1$; and the second uses $L_n(\theta,\beta_n;W_n)=0$, by Proposition~\ref{prop:weighted}(i) and $\theta\in\Theta_I$, together with the Lipschitz continuity of the extension, Proposition~\ref{prop:weighted}(iii).

\medskip\noindent\emph{Step 2: } By~(\ref{eq:relative}) at $\beta=\hat\beta_n$, on $G_n$,
\begin{eqnarray*}
\left|\hat T_n(\theta)-T_n\right| & \leq & (\kappa_n-1)\,\bar\ell_W\,r_n\|\hat\beta_n-\beta_n\| \; = \; (\kappa_n-1)\,\bar\ell_W\,\|Z_n\|,
\end{eqnarray*}
and the right-hand side converges to zero in probability, by~(\ref{eq:kappa-n}) and the uniform tightness of $(Z_n)$. Since $\mathbb P(G_n^c)\rightarrow0$, it follows that $\hat T_n(\theta)-T_n\rightarrow0$ in probability. As $|f(a)-f(b)|\leq\min\{2,|a-b|\}$ for every $f\in\mbox{BL}_1(\real)$ and all real $a,b$, we obtain $\sup_{f\in\mbox{\tiny BL}_1(\real)}|\mathbb Ef(\hat T_n(\theta))-\mathbb Ef(T_n)|\rightarrow0$, which combines with~(\ref{eq:rho-n}) for $T_n$ to give
\begin{eqnarray}
\label{eq:rho-n-hat}
\hat\rho_n & := & \sup_{f\in\mbox{\tiny BL}_1(\real)}\left|\mathbb E\left[f(\hat T_n(\theta))\right]-\mathbb E\left[f(L^\prime_n(\mathbb Z_0))\right]\right| \; \rightarrow \; 0.
\end{eqnarray}

\medskip\noindent\emph{Step 3: } For every $h\in\real^{d_\beta}$, by~(\ref{eq:relative}) at $\beta=\hat\beta_n+\epsilon_nh$ and at $\beta=\hat\beta_n$, on $G_n$,
\begin{eqnarray*}
\left|\hat L^\prime_n(h;\hat W_n)-\hat L^\prime_n(h)\right|
& \leq & \frac{(\kappa_n-1)\,\bar\ell_W}{\epsilon_n}\left( \|\hat\beta_n+\epsilon_nh-\beta_n\|+\|\hat\beta_n-\beta_n\|\right) \\
& \leq & (\kappa_n-1)\,\bar\ell_W\left( \frac{2\|Z_n\|}{r_n\epsilon_n}+\|h\|\right).
\end{eqnarray*}
Hence, for every $c>0$, $\sup_{\|h\|\leq c}|\hat L^\prime_n(h;\hat W_n)-\hat L^\prime_n(h)|\rightarrow0$ in probability, by~(\ref{eq:kappa-n}), $\mathbb P(G_n)\rightarrow1$, the tightness of $(Z_n)$ and $r_n\epsilon_n\rightarrow\infty$. Step~2 of the proof of Theorem~\ref{thm:ndm}, applied to the weighted binding function, gives $\sup_{\|h\|\leq c}|\hat L^\prime_n(h)-L^\prime_n(h)|\rightarrow0$ in probability, so that $\sup_{\|h\|\leq c}|\hat L^\prime_n(h;\hat W_n)-L^\prime_n(h)|\rightarrow0$ in probability for every $c>0$. The remainder of that step only uses this uniform approximation on balls, together with Assumption~\ref{ass:boot}, the Lipschitz continuity of $L^\prime_n$ and the uniform tightness in probability of $(\mathbb Z_n^\ast)$, and therefore yields
\begin{eqnarray}
\label{eq:rho-star-hat}
\hat\rho_n^\ast & := & \sup_{f\in\mbox{\tiny BL}_1(\real)}\left|\mathbb E\left[f(\hat L^\prime_n(\mathbb Z_n^\ast;\hat W_n))\,\vert\,(Y_i,X_i)_{i=1}^n\right]-\mathbb E\left[f(L^\prime_n(\mathbb Z_0))\right]\right| \; \rightarrow \; 0
\end{eqnarray}
in probability. No independence between $\hat W_n$ and $\mathbb Z_n^\ast$ is needed here: $\hat W_n$ and $\hat\beta_n$ are fixed conditionally on the data, and the supremum over $\|h\|\leq c$ is taken before the conditional expectation.

\medskip\noindent\emph{Conclusion.} In the nondegenerate case~(b) of Assumption~\ref{ass:cv} for the weight sequence, Step~3 of the proof of Theorem~\ref{thm:ndm} uses only the convergences $\rho_n\rightarrow0$ and $\rho_n^\ast\rightarrow0$ in probability, Assumption~\ref{ass:cv}(b) for the distribution function $F_n$ of $L^\prime_n(\mathbb Z_0)$, the definition of $\hat c_n$ as a conditional quantile of the statistic entering $\rho_n^\ast$, and the inequalities~(\ref{eq:blsmooth}). Applied with $\hat T_n(\theta)$, $\hat L^\prime_n(\mathbb Z_n^\ast;\hat W_n)$, $\hat\rho_n$ and $\hat\rho_n^\ast$ in place of $T_n$, $\hat L^\prime_n(\mathbb Z_n^\ast)$, $\rho_n$ and $\rho_n^\ast$, it yields $\limsup_n\mathbb P(\hat T_n(\theta)>\hat c_n)\leq\alpha$. In the degenerate case~(a), $L_n(\theta,\beta;W_n)=0$ for all $n\geq n_0$ and all $\beta$ with $\|\beta-\beta_n\|\leq\delta_0$; on $G_n$, where $\hat W_n$ is nonsingular, Proposition~\ref{prop:weighted}(i) gives $L_n(\theta,\beta;\hat W_n)=0$ for the same pairs $(n,\beta)$, so that the argument of the degenerate case in the proof of Theorem~\ref{thm:ndm} applies on $G_n$, with $\hat T_n(\theta)$ and $\hat L^\prime_n(\cdot;\hat W_n)$ in place of $T_n$ and $\hat L^\prime_n$, and gives $\mathbb P(\{\hat T_n(\theta)>\hat c_n\}\cap G_n)\rightarrow0$, while $\mathbb P(G_n^c)\rightarrow0$. In both cases, $\limsup_n\mathbb P(\hat T_n(\theta)>\hat c_n)\leq\alpha$ along the subsequence, conditionally on the covariates and almost surely, and Lemma~\ref{lem:transfer}(i) completes the proof.
\end{proof}

\section{Proofs of results in the main text}

\begin{proof}[Proof of Proposition~\ref{prop:idset}]
Under Assumption~\ref{ass:aux},
\begin{eqnarray*}
\frac1n\sum_{i=1}^n\mathbb E_{P_{Y\vert X,0}} \left[ s(Y\vert X_i;\beta_n)\; \vert\; X_i\right] & = & 0
\end{eqnarray*}
almost surely with respect to the distribution of $(X_i)_{i=1}^\infty$. If $\theta\in\Theta_I$, then $P_{Y\vert X,0}\in\mathcal P_\theta$. Hence there is at least one conditional distribution $P_{Y\vert X}$ in $\mathcal P_\theta$, namely $P_{Y\vert X,0}$ itself, such that
\begin{eqnarray*}
\frac1n\sum_{i=1}^n\mathbb E_{P_{Y\vert X}} \left[ s(Y\vert X_i;\beta_n)\; \vert\; X_i\right] & = & 0.
\end{eqnarray*}
Since $L_n(\theta,\beta_n)\geq0$ by definition, the result follows.
\end{proof}

\begin{proof}[Proof of Proposition~\ref{prop:dual}]
Fix $\theta$, $\beta$, $n$ and $X_1,\ldots,X_n$. Write $Q_x:=Q_{U\vert x;\theta_2}$ and $Q_i:=Q_{U\vert X_i;\theta_2}=Q_{X_i}$ for the conditional distribution of $U$ relative to $X_i$, and write $\Gamma_i(u):=\Gamma(u\vert X_i;\theta_1)$, $s_i(y):=s(y\vert X_i;\beta)$, $S_i:=s_i(Y)$ and $\Gamma^s_i(u):=s_i(\Gamma_i(u))\subseteq\real^{d_s}$. The proof proceeds in five steps: Steps~1--3 reduce the infinite dimensional program of Definition~\ref{def:bind-n} to a collection of per observation programs, Step~4 computes the value of the latter through support functions, and Step~5 concludes by minimax duality.

\medskip\noindent\emph{Step 1: } Under Assumption~\ref{ass:reg}(a)--(b), $\Gamma^s_i(u)$ is the image of a nonempty compact set under a continuous map, hence nonempty and compact, and $(u,x)\mapsto s(\Gamma(u\vert x;\theta_1)\vert x;\beta)$ is a jointly measurable correspondence, as the image of a jointly measurable compact valued correspondence under a Carath\'eodory function (see, e.g., \citealp{AB:99}); in particular $u\mapsto\Gamma^s_i(u)$ is measurable for each $i$. Under Assumption~\ref{ass:reg}(c), $\sup\{\|v\|:\;v\in\Gamma^s_i(u)\}\leq b(u,X_i;\theta)$ is $Q_i$-integrable, so that the random closed set ${\bf\Gamma}^s_i$ induced by $\Gamma^s_i$ and $Q_i$ is integrable, and $\mbox{Sel}^1(\Gamma^s_i,Q_i)=\mbox{Sel}^0(\Gamma^s_i)$ is nonempty by the Kuratowski--Ryll-Nardzewski selection theorem \citep[Theorem~18.13]{AB:99}.

\medskip\noindent\emph{Step 2: } Let
\begin{eqnarray*}
\mathcal M_i & := & \left\{ P \mbox{ on }\real^{d_s}:\;\exists\pi\in\mathcal M(P,Q_i)\mbox{ with }\pi\left( S\in\Gamma^s_i(U)\right)=1 \right\}
\end{eqnarray*}
denote the set of distributions of the auxiliary estimating function values compatible with the model at the covariate value $X_i$. We claim that $P\in\mathcal M_i$ if and only if there exist a distribution $\tilde P$ on $\mathcal Y$ and a coupling $\tilde\pi\in\mathcal M(\tilde P,Q_i)$ with $\tilde\pi(Y\in\Gamma_i(U))=1$ and $P=\tilde P\circ s_i^{-1}$; in particular, in that case $\mathbb E_P[S]=\mathbb E_{\tilde P}[s_i(Y)]$.

Sufficiency is immediate, since the pushforward of $\tilde\pi$ under $(y,u)\mapsto(s_i(y),u)$ is a coupling of $P$ and $Q_i$ concentrated on the graph of $\Gamma^s_i$. We now prove necessity. Let $\pi\in\mathcal M(P,Q_i)$ satisfy $\pi(S\in\Gamma^s_i(U))=1$ and define the correspondence
\begin{eqnarray*}
\Psi_i(v,u) & := & \left\{ y\in\Gamma_i(u):\; s_i(y)=v \right\},
\end{eqnarray*}
from $\real^{d_s}\times\mathcal U$ to $\mathcal Y$. Its values are compact, as closed subsets of the compact set $\Gamma_i(u)$ by continuity of $s_i$ (Assumption~\ref{ass:reg}(b)), and nonempty for $\pi$-almost every $(v,u)$ by definition of $\Gamma^s_i$. Its graph
\begin{eqnarray*}
\Gr\Psi_i & = & \left\{ (v,u,y):\; y\in\Gamma_i(u) \right\} \cap \left\{ (v,u,y):\; s_i(y)=v \right\}
\end{eqnarray*}
is measurable, as the intersection of the measurable graph of the compact valued measurable correspondence $\Gamma_i$ with a closed set. Hence $\Psi_i$ is a measurable, nonempty compact valued correspondence on a set of full $\pi$-measure, and the Kuratowski--Ryll-Nardzewski selection theorem yields a measurable map $g_i$ with $g_i(v,u)\in\Psi_i(v,u)$ for $\pi$-almost every $(v,u)$. Let $\tilde\pi$ be the pushforward of $\pi$ under $(v,u)\mapsto(g_i(v,u),u)$ and $\tilde P$ its $\mathcal Y$-marginal. By construction, $\tilde\pi$ has $\mathcal U$-marginal $Q_i$, satisfies $\tilde\pi(Y\in\Gamma_i(U))=1$, and $s_i(g_i(v,u))=v$ $\pi$-almost everywhere, so that $\tilde P\circ s_i^{-1}=P$, which proves the claim.

\medskip\noindent\emph{Step 3: } 
We claim that
\begin{eqnarray}
\label{eq:dual-percoord}
L_n(\theta,\beta) & = & \inf_{\forall i,\;P_i\in \mathcal M_i} \left\|\frac1n\sum_{i=1}^n\mathbb E_{P_i}S_i\right\|.
\end{eqnarray}
That the right-hand side of~(\ref{eq:dual-percoord}) is no larger than $L_n(\theta,\beta)$ is immediate: by Definition~\ref{def:DGP}, a kernel $P_{Y\vert X}\in\mathcal P_\theta$ induces, at each $X_i$, a distribution of $S_i$ that belongs to $\mathcal M_i$ by the sufficiency part of Step~2, and $\mathcal P_\theta\neq\varnothing$ by Assumption~\ref{ass:reg}(a), so the infimum on the left is over a nonempty set.

For the converse inequality, fix $(P_1,\ldots,P_n)\in\mathcal M_1\times\cdots\times\mathcal M_n$; we construct a kernel in $\mathcal P_\theta$ that attains the same value of the averaged estimating equations. By the necessity part of Step~2, each $P_i$ is associated with a coupling $\tilde\pi_i$ on $\mathcal Y\times\mathcal U$, with $\mathcal U$-marginal $Q_i$, concentrated on the graph of $\Gamma_i$, and whose $\mathcal Y$-marginal $\tilde P_i$ satisfies $\mathbb E_{\tilde P_i}[s(Y\vert X_i;\beta)]=\mathbb E_{P_i}S_i$. Let $x^{(1)},\ldots,x^{(m)}$ be the distinct values taken by $X_1,\ldots,X_n$ and let $I_g:=\{i\leq n:\;X_i=x^{(g)}\}$, so that the $I_g$ partition $\{1,\ldots,n\}$. Definition~\ref{def:DGP} forces a common conditional distribution across the observations in a tie group $I_g$, but this entails no loss. Indeed, the set $\mathcal C_g$ of couplings on $\mathcal Y\times\mathcal U$ with $\mathcal U$-marginal $Q_{x^{(g)}}$ that are concentrated on the graph of $\Gamma(\cdot\vert x^{(g)};\theta_1)$ is convex, since a mixture of couplings with a common marginal and a common support restriction satisfies both; hence
\begin{eqnarray*}
\bar\pi_g \; := \; \frac{1}{\vert I_g\vert}\sum_{i\in I_g}\tilde\pi_i \;\in\;\mathcal C_g,
& \quad &
\mathbb E_{\bar P_g}\left[ s(Y\vert x^{(g)};\beta)\right] \; = \; \frac{1}{\vert I_g\vert}\sum_{i\in I_g}\mathbb E_{P_i}S_i,
\end{eqnarray*}
where $\bar P_g$ is the $\mathcal Y$-marginal of $\bar\pi_g$. This is precisely the convexity property built into the convexified selection expectation: the mean attainable at a tied covariate value is the \emph{average} of the individual means, so that averaging over a tie group keeps the resulting vector in the feasible convex set, and~(\ref{eq:dual-percoord}) is unaffected by the tie.

It remains to assemble the $\bar\pi_g$, which are defined only at the finitely many observed covariate values, into a Markov kernel with source $\supp(X)$. By Assumption~\ref{ass:reg}(a) and the Kuratowski--Ryll-Nardzewski selection theorem, there is a jointly measurable selection $\gamma$ of $(u,x)\mapsto\Gamma(u\vert x;\theta_1)$; let $\pi^0_x$ denote the pushforward of $Q_x$ under $u\mapsto(\gamma(u,x),u)$, which is a Markov kernel with source $\supp(X)$, since $x\mapsto\int\mathbf 1\{\gamma(u,x)\in B\}\,Q_x(du)$ is measurable for every Borel $B$ by joint measurability of $\gamma$ and by $Q$ being a Markov kernel, and which satisfies the two restrictions of Definition~\ref{def:DGP} at every $x$. Now set
\begin{eqnarray*}
\pi_x & := & \left\{ \begin{array}{ll} \bar\pi_g & \mbox{if } x=x^{(g)},\; 1\leq g\leq m,\\[4pt] \pi^0_x & \mbox{otherwise.}\end{array}\right.
\end{eqnarray*}
Since $\{x^{(1)},\ldots,x^{(m)}\}$ is a finite, hence Borel, subset of $\supp(X)$, the map $x\mapsto\pi_x$ remains a Markov kernel, and it satisfies the support and marginal restrictions of Definition~\ref{def:DGP} at every $x\in\supp(X)$. Its $\mathcal Y$-marginal $P_{Y\vert x}$ therefore belongs to $\mathcal P_\theta$ and satisfies
\begin{eqnarray*}
\frac1n\sum_{i=1}^n\mathbb E_{P_{Y\vert X_i}}\left[ s(Y\vert X_i;\beta)\right]
\; = \; \frac1n\sum_{g=1}^m \vert I_g\vert\, \mathbb E_{\bar P_g}\left[ s(Y\vert x^{(g)};\beta)\right]
\; = \; \frac1n\sum_{i=1}^n\mathbb E_{P_i}S_i,
\end{eqnarray*}
which establishes~(\ref{eq:dual-percoord}).

\medskip\noindent\emph{Step 4: } 
In view of~(\ref{eq:dual-percoord}), and recalling that $h(a,A):=\inf_{y\in A}a^\top y$ denotes the (lower) support function of the set $A$ in direction $a$, it suffices to prove that
\begin{eqnarray}
\label{eq:dual-proof}
\sup_{\lambda\in\mathcal S} \left( \frac1n\sum_{i=1}^n \mathbb E_{Q_i} \left[ \inf_{s_i\in\Gamma^s_i(U)} \lambda^\top s_i \right] \right)
& = &
\inf_{\forall i,\;P_i\in \mathcal M_i} \left\|\frac1n\sum_{i=1}^n\mathbb E_{P_i}S_i\right\|.
\end{eqnarray}
The left-hand side of~(\ref{eq:dual-proof}) is equal to $\sup_{\lambda\in \mathcal S}\frac1n\sum_{i=1}^n\mathbb E \,h(\lambda,{\bf\Gamma}^s_i)$, the measurability of $u\mapsto h(\lambda,\Gamma^s_i(u))$ following from the measurable maximum theorem (see, e.g., \citealp{AB:99}). By the Hiai--Umegaki interchange of infimum and integral for integrable random closed sets \citep[Theorem~2.2]{HU:77}, applicable by Step~1,
\begin{eqnarray*}
\mathbb E\, h(\lambda,{\bf\Gamma}^s_i)
\; = \; \mathbb E_{Q_i}\left[ \inf_{v\in\Gamma^s_i(U)}\lambda^\top v\right]
\; = \; \inf_{\gamma\in\text{\normalfont Sel}^1(\Gamma^s_i,Q_i)} \lambda^\top\,\mathbb E_{Q_i}\left[\gamma(U)\right]
\; = \; h\left(\lambda,{\bf E}{\bf\Gamma}^s_i\right),
\end{eqnarray*}
the last equality because a lower support function is invariant under the closed convex hull of its argument, so that the value of $h(\lambda,\cdot)$ at the Aumann expectation and at the convexified selection expectation coincide; see also Theorem 2.1.35 page 247 of \cite{Molchanov:2017}. 
Hence
\[
\frac1n\sum_{i=1}^n\mathbb E\, h(\lambda,{\bf\Gamma}^s_i)=\frac1n\sum_{i=1}^nh(\lambda,{\bf E} {\bf\Gamma}^s_i)=h\left( \lambda,\frac1n\bigoplus_{i=1}^n{\bf E} {\bf\Gamma}^s_i \right),
\]
where $\bigoplus$ denotes Minkowski addition and we used the positive homogeneity and additivity of support functions under scaling and Minkowski sums.

\medskip\noindent\emph{Step 5: } Write $\mathcal K_n(\theta,\beta):=\frac1n\bigoplus_{i=1}^n{\bf E}{\bf\Gamma}^s_i$. Each ${\bf E}{\bf\Gamma}^s_i$ is convex by construction, and compact, being a closed and bounded subset of $\real^{d_s}$, the bound $\|\mathbb E_{Q_i}\gamma(U)\|\leq\mathbb E_{Q_i}[b(U,X_i;\theta)]<\infty$ holding uniformly over selections by Assumption~\ref{ass:reg}(c); hence $\mathcal K_n(\theta,\beta)$ is compact and convex. By Step~4, the left-hand side of~(\ref{eq:dual-proof}) is equal to
\begin{eqnarray}
\label{eq:dual-proof2}
\sup_{\lambda\in \mathcal S}h\left(\lambda,\mathcal K_n(\theta,\beta)\right)=\inf_{\mu\in \mathcal K_n(\theta,\beta)}\|\mu\| \; = \; \inf_{\forall i,\;\mu_i\in{\bf E}{\bf\Gamma}^s_i}\left\|\frac1n\sum_{i=1}^n\mu_i\right\|,
\end{eqnarray}
where the first equality holds because, for any compact convex set $C$ and $\mathcal S$ the closed unit ball, the following minimax result holds by \cite{Sion:58}:
\[
\sup_{\lambda\in \mathcal S}\inf_{\mu\in C}\lambda^\top\mu=\inf_{\mu\in C}\sup_{\lambda\in \mathcal S}\lambda^\top\mu=\inf_{\mu\in C}\|\mu\|.
\]
Finally, the right-hand side of~(\ref{eq:dual-proof2}) is equal to the right-hand side of~(\ref{eq:dual-proof}) by Lemma~\ref{lem:Strass}, applied to each ${\bf\Gamma}^s_i$ separately.
\end{proof}

\begin{lemma}
\label{lem:Strass}
Let ${\bf\Gamma}$ be the random closed set induced by a measurable, nonempty compact valued correspondence $\Gamma$ from $\mathcal U$ to $\real^d$ and a distribution $Q$, and suppose that $b(u):=\sup\{\|v\|:\;v\in\Gamma(u)\}$ is $Q$-integrable. Let
$
\mathcal M_\Gamma := \left\{P:\;\exists \pi\in\mathcal M(P,Q):\;\pi(Y\in\Gamma(U))=1\right\}.
$
Then, with ${\bf E}{\bf\Gamma}$ the convexified selection expectation of the notation subsection,
$
\left\{ \mathbb E_PY:\;P\in\mathcal M_\Gamma\right\} = {\bf E}{\bf\Gamma}$, and in particular
$
\inf_{\mu\in{\bf E}{\bf\Gamma}}\|\mu\|=\inf \left\{\left\|\mathbb E_{P}Y\right\|:\;P\in\mathcal M_\Gamma\right\}.
$
\end{lemma}

\begin{proof}[Proof of Lemma~\ref{lem:Strass}]
By Proposition~1 of \cite{EGH:2010}, $P\in\mathcal M_\Gamma$ if and only if $P$ satisfies the Artstein inequalities (see \citealp{Molchanov:2017}, page~82), i.e., if and only if $P(B)\leq Q(\{u:\Gamma(u)\cap B\ne\varnothing\})$ for all Borel measurable $B$. Every $P\in\mathcal M_\Gamma$ is supported, through its coupling with $Q$, on values dominated in norm by $b(U)$, so that $P(\|Y\|>K)\leq Q(b(U)>K)$ and the family $\mathcal M_\Gamma$ is both uniformly tight and uniformly integrable:
\begin{eqnarray*}
\sup_{P\in\mathcal M_\Gamma}\mathbb E_P\left[\|Y\|\mathbf{1}\{\|Y\|>K\}\right]\;\leq\;\mathbb E_Q\left[b(U)\mathbf{1}\{b(U)>K\}\right]\;\longrightarrow\;0
\end{eqnarray*}
as $K\rightarrow\infty$. Write $A:=\{\mathbb E_Q[\gamma(U)]:\;\gamma\in\mbox{Sel}^1(\Gamma,Q)\}$ for the Aumann expectation, so that ${\bf E}{\bf\Gamma}=\clco A$, a compact convex set since $A$ is bounded by $\mathbb E_Q[b(U)]$.

We show that $\{\mathbb E_PY:\;P\in\mathcal M_\Gamma\}=\clco A$, from which the displayed identity follows by taking infima of the norm over both sets. For the inclusion $\supseteq$, note first that $\mathcal M_\Gamma$ is convex, since mixtures of couplings concentrated on the graph of $\Gamma$ with common marginal $Q$ remain so concentrated with the same marginal. Every element of $\co A$ is the mean of a finite mixture of laws $Q\gamma^{-1}$ of measurable selections, each of which belongs to $\mathcal M_\Gamma$, and hence is the mean of an element of $\mathcal M_\Gamma$ by convexity; thus $\co A\subseteq\{\mathbb E_PY:\;P\in\mathcal M_\Gamma\}$. The latter set is closed: if $\mathbb E_{P_k}Y\rightarrow\mu$ with $P_k\in\mathcal M_\Gamma$, uniform tightness gives a subsequence along which $P_k\weakconv P$, and the Artstein inequalities pass to the weak limit. 
Uniform integrability then gives $\mathbb E_{P_k}Y\rightarrow\mathbb E_PY=\mu$. Hence $\clco A\subseteq\{\mathbb E_PY:\;P\in\mathcal M_\Gamma\}$.

For the reverse inclusion, let $P\in\mathcal M_\Gamma$. By Corollary~3.4 of \cite{CMM:2004}, $\mathcal M_\Gamma$ is equal to the weak closure (in the sense of convergence in distribution) of the convex hull of the set $\{P:\;P=Q\gamma^{-1},\; \gamma\in\mbox{Sel}^0(\Gamma)\}$, so that $P$ is the weak limit of a sequence $(P_k)$ of finite mixtures of laws of measurable selections. The mean of each $P_k$ is a convex combination of selection expectations, hence an element of $\co A$, and $\mathbb E_{P_k}Y\rightarrow\mathbb E_PY$ by weak convergence combined with the uniform integrability established above. Since $\clco A$ is closed, $\mathbb E_PY\in\clco A$. The two inclusions give the announced equality.
\end{proof}

\begin{proof}[Proof of Proposition~\ref{prop:hdd}]
Fix $\theta\in\Theta$ and $n\geq1$, and work on the almost sure set of covariate sequences on which Assumptions~\ref{ass:aux} and~\ref{ass:uniq} hold and on which $\frac1n\sum_{i=1}^n\mathbb E_{Q_i}[b(U,X_i;\theta)]\rightarrow\mathbb E_{P_{X,0}}\mathbb E_{Q_{U\vert X;\theta_2}}[b(U,X;\theta)]<\infty$, the latter by the strong law of large numbers under Assumption~\ref{ass:reg}(c). Write $Q_i:=Q_{U\vert X_i;\theta_2}$, $\Gamma_i(u):=\Gamma(u\vert X_i;\theta_1)$, $s_i(y,\beta):=s(y\vert X_i;\beta)$ and $b_i(u):=b(u,X_i;\theta)$, and recall the dual objective $\varphi_n(\lambda,\beta)$ of~(\ref{eq:phin}), the inner value function
\begin{eqnarray*}
v_i(u,\lambda,\beta) & := & \min_{y\in\Gamma_i(u)}\lambda^\top s_i(y,\beta),
\end{eqnarray*}
whose minimum is attained by Assumption~\ref{ass:reg}(a)--(b), and the solution sets $\hat{\mathcal Y}_i(u,\lambda,\beta)$ and $\hat\Lambda_n(\theta,\beta)$, so that $\varphi_n(\lambda,\beta)=\frac1n\sum_{i=1}^n\mathbb E_{Q_i}[v_i(U,\lambda,\beta)]$ and, by Proposition~\ref{prop:dual}, we have $L_n(\theta,\beta)=\sup_{\lambda\in\mathcal S}\varphi_n(\lambda,\beta)$. Measurability of $u\mapsto v_i(u,\lambda,\beta)$ and the existence of measurable selections from $\hat{\mathcal Y}_i(\cdot,\lambda,\beta)$ follow from the measurable maximum theorem (see, e.g., \citealp{AB:99}). We proceed in five steps.

\medskip\noindent\emph{Step 1: } For every $\lambda\in\mathcal S$, $u\in\mathcal U$, $1\leq i\leq n$ and $\beta,\beta^\prime\in\mathbb B_0$,
\begin{eqnarray*}
\left| v_i(u,\lambda,\beta) - v_i(u,\lambda,\beta^\prime) \right|
& \leq & \sup_{y\in\Gamma_i(u)}\left| \lambda^\top\left( s_i(y,\beta)-s_i(y,\beta^\prime)\right)\right|
\; \leq \; b_i(u)\,\|\beta-\beta^\prime\|,
\end{eqnarray*}
where the first inequality uses $\|\lambda\|\leq1$ and the fact that the difference of infima is bounded by the supremum of the differences, and the second follows from the mean value theorem applied to $t\mapsto\lambda^\top s_i(y,\beta^\prime+t(\beta-\beta^\prime))$ on $[0,1]$. The latter can be applied, since $\mathbb B_0$ is convex and $\|\nabla_\beta s_i(y,\cdot)\|\leq b_i(u)$ on $\Gamma_i(u)\times\mathbb B_0$ by Assumption~\ref{ass:reg}(b)--(c). Taking $Q_i$-expectations, averaging over $i$, and taking the supremum over $\lambda\in\mathcal S$, which preserves the Lipschitz property, yields
\begin{eqnarray*}
\left| L_n(\theta,\beta)-L_n(\theta,\beta^\prime)\right| & \leq & \ell_n\,\|\beta-\beta^\prime\|,
\qquad \ell_n:=\frac1n\sum_{i=1}^n\mathbb E_{Q_i}\left[ b_i(U)\right],
\end{eqnarray*}
and $\sup_n\ell_n<\infty$ on the almost sure set fixed at the outset, since $\ell_n$ converges. The same argument before taking the supremum over $\lambda$ shows that $\varphi_n(\cdot,\beta)$ is $\ell_n$-Lipschitz in $\beta$ uniformly in $\lambda\in\mathcal S$, and the bound $|\varphi_n(\lambda,\beta)-\varphi_n(\lambda^\prime,\beta)|\leq\|\lambda-\lambda^\prime\|\ell_n$, obtained analogously, shows that $\varphi_n$ is jointly continuous on $\mathcal S\times\mathbb B_0$; in particular $\hat\Lambda_n(\theta,\beta)$ is nonempty and compact for every $\beta$, and the correspondence $\beta\mapsto\hat\Lambda_n(\theta,\beta)$ is upper hemicontinuous with compact values by Berge's maximum theorem (see, e.g., \citealp{AB:99}). This proves part~(i).

\medskip\noindent\emph{Step 2: } Take $n$ sufficiently large that the ball of radius $\delta_\ast$ around $\beta_n$ lies in $\mathbb B_0$, as required by Assumption~\ref{ass:reg}(d), and fix $u$, $i$ and $\lambda$. We apply Proposition~4.12 page 272 of \cite{BS:2000} to the parameterized problem $\min_{y\in\Gamma_i(u)}f(y,\beta)$ with $f(y,\beta):=\lambda^\top s_i(y,\beta)$, at the point $\beta_n$. Condition~(1) of that proposition, continuity of $f$ on $\Gamma_i(u)\times\mathbb B_0$, holds by Assumption~\ref{ass:reg}(b). Condition~(2), inf-compactness, holds with $C=\Gamma_i(u)$, which is compact by Assumption~\ref{ass:reg}(a). Condition~(3), Hadamard directional differentiability of $\beta\mapsto f(y,\beta)$ at $\beta_n$ for each fixed $y$, holds since $f(y,\cdot)$ is continuously differentiable, with derivative $f^\prime_y(\beta_n,d)=\lambda^\top\nabla_\beta s_i(y,\beta_n)d$. For condition~(4), let $t_k\downarrow0$, $d_k\rightarrow d$, and let $(y_k)$ be any sequence in $\Gamma_i(u)$; by compactness it has a limit point $\bar y\in\Gamma_i(u)$, and along the corresponding subsequence, the mean value theorem gives
\begin{eqnarray*}
\frac{f(y_k,\beta_n+t_kd_k)-f(y_k,\beta_n)}{t_k} & = & \lambda^\top\nabla_\beta s_i(y_k,\tilde\beta_k)\,d_k
\end{eqnarray*}
for some $\tilde\beta_k$ on the segment joining $\beta_n$ and $\beta_n+t_kd_k$, and the right-hand side converges to $\lambda^\top\nabla_\beta s_i(\bar y,\beta_n)d=f^\prime_{\bar y}(\beta_n,d)$ by the joint continuity of $\nabla_\beta s_i$ (Assumption~\ref{ass:reg}(b)), so that condition~(4) holds with equality. Proposition~4.12 of \cite{BS:2000} then delivers Hadamard directional differentiability of $\beta\mapsto v_i(u,\lambda,\beta)$ at $\beta_n$, with derivative
\begin{eqnarray}
\label{eq:innerderiv}
v^\prime_i(u,\lambda,\beta_n;d) & = & \min_{y\in\hat{\mathcal Y}_i(u,\lambda,\beta_n)}\lambda^\top\nabla_\beta s_i(y,\beta_n)\,d.
\end{eqnarray}

\medskip\noindent\emph{Step 3: } Fix $\lambda\in\mathcal S$, $t_k\downarrow0$ and $d_k\rightarrow d$. By Step~1, the difference quotients 
\begin{eqnarray*}
    \left(v_i(u,\lambda,\beta_n+t_kd_k)-v_i(u,\lambda,\beta_n)\right)/t_k
\end{eqnarray*} are bounded in absolute value by $b_i(u)\sup_k\|d_k\|$, which is $Q_i$-integrable by Assumption~\ref{ass:reg}(c), and by Step~2 they converge pointwise in $u$ to $v^\prime_i(u,\lambda,\beta_n;d)$. Dominated convergence therefore yields Hadamard directional differentiability of $\beta\mapsto\varphi_n(\lambda,\beta)$ at $\beta_n$, with derivative
\begin{eqnarray}
\label{eq:phideriv}
\varphi^\prime_{n,\lambda}(\beta_n;d) & = & \frac1n\sum_{i=1}^n\mathbb E_{Q_i}\left[ \min_{y\in\hat{\mathcal Y}_i(U,\lambda,\beta_n)}\lambda^\top\nabla_\beta s_i(y,\beta_n)\,d\right].
\end{eqnarray}

\medskip\noindent\emph{Step 4: } Fix a direction $d$ and let $t\downarrow0$; write $V(\beta):=L_n(\theta,\beta)=\max_{\lambda\in\mathcal S}\varphi_n(\lambda,\beta)$, the maximum being attained by continuity of $\varphi_n(\cdot,\beta)$ on the compact set $\mathcal S$.

\emph{Lower bound.} For any $\lambda\in\hat\Lambda_n(\theta,\beta_n)$, $V(\beta_n+td)\geq\varphi_n(\lambda,\beta_n+td)$ and $V(\beta_n)=\varphi_n(\lambda,\beta_n)$, so that, by Step~3,
\begin{eqnarray}
\label{eq:lower}
\liminf_{t\downarrow0}\frac{V(\beta_n+td)-V(\beta_n)}{t} & \geq & \sup_{\lambda\in\hat\Lambda_n(\theta,\beta_n)}\varphi^\prime_{n,\lambda}(\beta_n;d).
\end{eqnarray}

\emph{Upper bound.} Let $(t_k)$ be a sequence achieving the $\limsup$ of the difference quotient of $V$, and pick $\lambda_k\in\hat\Lambda_n(\theta,\beta_n+t_kd)$, so that
\begin{eqnarray*}
\frac{V(\beta_n+t_kd)-V(\beta_n)}{t_k} & \leq & \frac{\varphi_n(\lambda_k,\beta_n+t_kd)-\varphi_n(\lambda_k,\beta_n)}{t_k} \;\; =: \;\; q_k.
\end{eqnarray*}
Passing to a further subsequence, $\lambda_k\rightarrow\bar\lambda$ for some $\bar\lambda\in\mathcal S$ by compactness, and $\bar\lambda\in\hat\Lambda_n(\theta,\beta_n)$ by the upper hemicontinuity of $\beta\mapsto\hat\Lambda_n(\theta,\beta)$ established in Step~1. We distinguish two cases.

If $\bar\lambda=0$, then, by the argument of Step~1, $|q_k|\leq\|\lambda_k\|\,\ell_n\|d\|\rightarrow0=\varphi^\prime_{n,0}(\beta_n;d)$, since $\varphi_n(0,\cdot)\equiv0$.

If $\bar\lambda\neq0$, fix $i$ and $u$ outside the $Q_i$-null set of Assumption~\ref{ass:uniq}, and write $g_i(u,\bar\lambda)$ for the common value of $\bar\lambda^\top\nabla_\beta s_i(y,\beta_n)$ over $y\in\hat{\mathcal Y}_i(u,\bar\lambda,\beta_n)$. Pick $y_k\in\hat{\mathcal Y}_i(u,\lambda_k,\beta_n)$; then
\begin{eqnarray*}
\frac{v_i(u,\lambda_k,\beta_n+t_kd)-v_i(u,\lambda_k,\beta_n)}{t_k}
& \leq & \frac{\lambda_k^\top\left( s_i(y_k,\beta_n+t_kd)-s_i(y_k,\beta_n)\right)}{t_k}
\; = \; \lambda_k^\top\nabla_\beta s_i(y_k,\tilde\beta_k)\,d,
\end{eqnarray*}
using the suboptimality of $y_k$ in the perturbed problem for the inequality and the mean value theorem for the equality. By compactness of $\Gamma_i(u)$, every subsequence of $(y_k)$ has a convergent further subsequence. Its limit belongs to $\hat{\mathcal Y}_i(u,\bar\lambda,\beta_n)$ by upper hemicontinuity of the argmin correspondence. Since $\lambda_k\rightarrow\bar\lambda$ and $\tilde\beta_k\rightarrow\beta_n$, joint continuity of $\nabla_\beta s_i$ and Assumption~\ref{ass:uniq} imply that the projected gradients converge to $g_i(u,\bar\lambda)$ along every such further subsequence. Hence $\lambda_k^\top\nabla_\beta s_i(y_k,\tilde\beta_k)\rightarrow g_i(u,\bar\lambda)$ along the full sequence, and
\begin{eqnarray*}
\limsup_{k}\;\frac{v_i(u,\lambda_k,\beta_n+t_kd)-v_i(u,\lambda_k,\beta_n)}{t_k}
& \leq & g_i(u,\bar\lambda)\,d \\
& = & \min_{y\in\hat{\mathcal Y}_i(u,\bar\lambda,\beta_n)}\bar\lambda^\top\nabla_\beta s_i(y,\beta_n)\,d,
\end{eqnarray*}
the last equality following from the definition of $g_i(u,\bar\lambda)$. Since the difference quotients are dominated by the $Q_i$-integrable envelope $b_i(u)\|d\|$ (Step~1), the reverse Fatou lemma yields $\limsup_kq_k\leq\varphi^\prime_{n,\bar\lambda}(\beta_n;d)$, with $\varphi^\prime_{n,\bar\lambda}$ as in~(\ref{eq:phideriv}).

In either case, $\limsup_{t\downarrow0}\left( V(\beta_n+td)-V(\beta_n)\right)/t\leq\varphi^\prime_{n,\bar\lambda}(\beta_n;d)$ for some $\bar\lambda\in\hat\Lambda_n(\theta,\beta_n)$, which combines with the lower bound~(\ref{eq:lower}) to give
\begin{eqnarray*}
\lim_{t\downarrow0}\frac{V(\beta_n+td)-V(\beta_n)}{t} & = & \sup_{\lambda\in\hat\Lambda_n(\theta,\beta_n)}\varphi^\prime_{n,\lambda}(\beta_n;d) \; = \; \varphi^\prime_{n,\bar\lambda}(\beta_n;d),
\end{eqnarray*}
so that the supremum is attained at $\bar\lambda$ and the limit exists, i.e., $\beta\mapsto L_n(\theta,\beta)$ is directionally differentiable at $\beta_n$ in the ordinary sense, with derivative given by~(\ref{eq:hdd}).

\medskip\noindent\emph{Step 5: } By Step~1, $\beta\mapsto L_n(\theta,\beta)$ is Lipschitz on $\mathbb B_0$; for locally Lipschitz maps, directional differentiability in the ordinary sense implies Hadamard directional differentiability, with the same derivative (see \citealp{Shapiro:90}). This proves part~(ii) and completes the proof.
\end{proof}

The proofs of Theorems~\ref{thm:ndm} and~\ref{thm:weights} condition on the covariate sequence. The following lemma transfers statements that hold conditionally on the covariates along subsequences to the statements of the theorems, and converts the convergence in probability of Assumptions~\ref{ass:boot} and~\ref{ass:weights} into convergence in probability conditionally on the covariates along subsequences. Throughout, $\mathbf X:=(X_i)_{i=1}^\infty$.

\begin{lemma}
\label{lem:transfer}
\begin{enumerate}[label=(\roman*)]
\item Let $(E_n)_{n\geq1}$ be events and $\alpha\in[0,1]$. If every subsequence of $(n)_{n\geq1}$ has a further subsequence $(n_j)_{j\geq1}$ along which $\limsup_j\mathbb P(E_{n_j}\,\vert\,\mathbf X)\leq\alpha$ almost surely, then $\limsup_n\mathbb P(E_n)\leq\alpha$.
\item Let $(V_n)_{n\geq1}$ be random variables with $V_n\rightarrow0$ in probability. Then every subsequence of $(n)_{n\geq1}$ has a further subsequence $(n_j)_{j\geq1}$ along which $V_{n_j}\rightarrow0$ in probability conditionally on $\mathbf X$, almost surely with respect to the distribution of $\mathbf X$.
\end{enumerate}
\end{lemma}

\begin{proof}[Proof of Lemma~\ref{lem:transfer}]
(i) Let $(n_k)$ be a subsequence along which $\mathbb P(E_{n_k})\rightarrow\limsup_n\mathbb P(E_n)$, and $(n_{k_j})$ a further subsequence as in the hypothesis. Since $0\leq\mathbb P(E_{n_{k_j}}\vert\mathbf X)\leq1$, the reverse Fatou lemma gives
\begin{eqnarray*}
\limsup_n\mathbb P(E_n) \; = \; \lim_j\mathbb E\left[\mathbb P(E_{n_{k_j}}\,\vert\,\mathbf X)\right] \; \leq \; \mathbb E\left[\limsup_j\mathbb P(E_{n_{k_j}}\,\vert\,\mathbf X)\right] \; \leq \; \alpha.
\end{eqnarray*}
(ii) Every subsequence of a sequence converging to zero in probability has a further subsequence $(n_j)$ along which $V_{n_j}\rightarrow0$ almost surely. Fix $\epsilon>0$ and let $N_\epsilon:=\limsup_j\{|V_{n_j}|>\epsilon\}$, an event of probability zero, so that $\mathbb P(N_\epsilon\vert\mathbf X)=0$ almost surely. By the conditional version of the reverse Fatou lemma, $\limsup_j\mathbb P(|V_{n_j}|>\epsilon\,\vert\,\mathbf X)\leq\mathbb P(N_\epsilon\vert\mathbf X)=0$ almost surely. Taking $\epsilon$ along a countable sequence decreasing to zero gives the claim on a single almost sure set.
\end{proof}

\begin{proof}[Proof of Theorem~\ref{thm:ndm}]
Fix $\theta\in\Theta_I$ and $\alpha\in(0,1)$, write $E_n:=\{r_nL_n(\theta,\hat\beta_n)>\hat c_{n,1-\alpha}(\theta)\}$ for the rejection event, and let $\rho_n^{\mathbb Z}$ denote the left-hand side of Assumption~\ref{ass:boot}, a measurable function of the data.

\medskip\noindent\emph{Reduction to a conditional statement.} By Lemma~\ref{lem:transfer}(i), it suffices to show that every subsequence of $(n)_{n\geq1}$ has a further subsequence along which $\limsup_n\mathbb P(E_n\,\vert\,(X_i)_{i=1}^\infty)\leq\alpha$ almost surely. Since $\rho_n^{\mathbb Z}\rightarrow0$ in probability, Lemma~\ref{lem:transfer}(ii) provides, within every subsequence, a further subsequence along which $\rho_n^{\mathbb Z}\rightarrow0$ in probability conditionally on $(X_i)_{i=1}^\infty$, almost surely with respect to the distribution of $(X_i)_{i=1}^\infty$. Assumptions~\ref{ass:est}, \ref{ass:reg}(d), \ref{ass:unif} and~\ref{ass:cv} are limit statements along the full sequence, and hold along every subsequence, since passing to a subsequence can only decrease a $\limsup$ and increase a $\liminf$; the remaining assumptions and the conclusions of Propositions~\ref{prop:idset} and~\ref{prop:hdd} are statements about each $n$. We therefore fix such a subsequence, along which all limits below are taken and whose elements we still denote by $n$, and condition throughout on a realization of $(X_i)_{i=1}^\infty$ in the almost sure set on which Assumptions~\ref{ass:aux}, \ref{ass:est}, \ref{ass:reg}--\ref{ass:cv} and the conclusions of Propositions~\ref{prop:idset} and~\ref{prop:hdd} hold and on which $\rho_n^{\mathbb Z}\rightarrow0$ in probability conditionally on the covariates along the subsequence; all probability statements below are conditional on the covariates, with randomness coming from $(Y_i)_{i=1}^n$ and the resampling scheme, and $\rho_n^{\mathbb Z}\rightarrow0$ in probability in this conditional sense. Write $Z_n:=r_n(\hat\beta_n-\beta_n)$, $T_n:=r_nL_n(\theta,\hat\beta_n)$, $L^\prime_n(h):=L^\prime_{\beta_n}(\theta,h)$, $c_n:=c_{n,1-\alpha}(\theta)$, $\hat c_n:=\hat c_{n,1-\alpha}(\theta)$, and $\bar\ell:=\sup_n\ell_n<\infty$, with $\ell_n$ the Lipschitz constant of Proposition~\ref{prop:hdd}(i); note that $h\mapsto L^\prime_n(h)$ is $\ell_n$-Lipschitz, as a pointwise limit of $\ell_n$-Lipschitz difference quotients. We record two facts used repeatedly. First, $(Z_n)$ is uniformly tight: by Assumption~\ref{ass:est} and tightness of $\mathbb Z_0$, for every $\eta>0$ there is $c>0$ with $\limsup_n\mathbb P(\|Z_n\|>c)\leq\eta$, as is seen by testing against the bounded Lipschitz function $h\mapsto\min\{1,\max\{0,\|h\|-c+1\}\}$; the same argument applied to $\rho_n^{\mathbb Z}\rightarrow0$ shows that $(\mathbb Z_n^\ast)$ is uniformly tight in probability. Second, for random variables $A$, $B$ with $\rho:=\sup_{f\in\mbox{\tiny BL}_1(\mathbb R)}|\mathbb E f(A)-\mathbb E f(B)|$ and any $x\in\real$, $\delta>0$, we have
\begin{eqnarray}
\label{eq:blsmooth}
\mathbb P(A>x)\;\leq\;\mathbb P(B>x-\delta)+(1+1/\delta)\rho, \;
\mathbb P(A\leq x)\;\leq\;\mathbb P(B\leq x+\delta)+(1+1/\delta)\rho.
\end{eqnarray}

\medskip\noindent\emph{Degenerate case.} Suppose first that case~(a) of Assumption~\ref{ass:cv} holds, with associated $n_0$ and $\delta_0$. Since $\|\hat\beta_n-\beta_n\|=\|Z_n\|/r_n\rightarrow0$ in probability by tightness of $(Z_n)$ and $r_n\rightarrow\infty$, the event $A_n:=\{\|\hat\beta_n-\beta_n\|\leq\delta_0/2\}$ satisfies $\mathbb P(A_n)\rightarrow1$. On $A_n$, for $n\geq n_0$, $L_n(\theta,\hat\beta_n)=0$, hence $T_n=0$; moreover $\hat L^\prime_{\beta_n}(\theta,h)=L_n(\theta,\hat\beta_n+\epsilon_nh)/\epsilon_n\geq0$ for every $h\in\real^{d_\beta}$, by nonnegativity of the extension, and $\hat L^\prime_{\beta_n}(\theta,\mathbb Z_n^\ast)=0$ whenever $\epsilon_n\|\mathbb Z_n^\ast\|\leq\delta_0/2$, an event whose conditional probability given the data exceeds $1-\alpha$ with probability tending to one, by uniform tightness in probability of $(\mathbb Z_n^\ast)$ and $\epsilon_n\rightarrow0$. On the intersection of these events, $\hat c_n=0$ and $T_n=0$, so that $\mathbb P(T_n>\hat c_n)\rightarrow0\leq\alpha$.

\medskip\noindent\emph{Nondegenerate case.} Suppose now that case~(b) of Assumption~\ref{ass:cv} holds. We proceed in three steps.

\medskip\noindent\emph{Step 1: } By Proposition~\ref{prop:idset}, $L_n(\theta,\beta_n)=0$, so that $T_n=r_n\left( L_n(\theta,\hat\beta_n)-L_n(\theta,\beta_n)\right)$. Fix $c>0$; on the event $\{\|Z_n\|\leq c\}$,
\begin{eqnarray*}
\left| T_n - L^\prime_n(Z_n) \right| & \leq & a_n(c) \; := \; \sup_{\|h\|\leq c}\left| r_n\left( L_n(\theta,\beta_n+h/r_n)-L_n(\theta,\beta_n)\right) - L^\prime_n(h) \right| \; \rightarrow \; 0,
\end{eqnarray*}
by Assumption~\ref{ass:unif} applied with $t_n=1/r_n$, using the positive homogeneity of $L^\prime_n$. Hence, for any $f\in\mbox{BL}_1(\real)$,
\begin{eqnarray*}
\left|\mathbb E f(T_n)-\mathbb E f(L^\prime_n(\mathbb Z_0))\right|
& \leq & a_n(c) + 2\,\mathbb P(\|Z_n\|>c) + \left|\mathbb E f(L^\prime_n(Z_n))-\mathbb E f(L^\prime_n(\mathbb Z_0))\right|,
\end{eqnarray*}
and the last term is bounded by $\max\{\bar\ell,1\}$ times the bounded Lipschitz distance in Assumption~\ref{ass:est}, since $f\circ L^\prime_n$ is bounded by $1$ and Lipschitz with constant at most $\bar\ell$. Taking the supremum over $f$, then $\limsup_n$, then letting $c\rightarrow\infty$ along the uniform tightness of $(Z_n)$, we conclude that
\begin{eqnarray}
\label{eq:rho-n}
\rho_n & := & \sup_{f\in\mbox{\tiny BL}_1(\mathbb R)}\left| \mathbb E\left[ f(T_n)\right]
-\mathbb E\left[ f(L^\prime_n(\mathbb Z_0)) \right] \right| \; \rightarrow \; 0.
\end{eqnarray}

\medskip\noindent\emph{Step 2: } For any $h\in\real^{d_\beta}$, the Lipschitz property of $L_n(\theta,\cdot)$ gives
\begin{eqnarray*}
\left| \hat L^\prime_{\beta_n}(\theta,h) - \frac{L_n(\theta,\beta_n+\epsilon_nh)-L_n(\theta,\beta_n)}{\epsilon_n} \right|
& \leq & \frac{2\ell_n\|\hat\beta_n-\beta_n\|}{\epsilon_n} \; = \; \frac{2\ell_n\|Z_n\|}{r_n\epsilon_n},
\end{eqnarray*}
uniformly in $h\in\real^{d_\beta}$, the two evaluations being well defined for every such $h$, and the bound holding with a constant not depending on $h$, because $L_n(\theta,\cdot)$ is understood as the globally $\ell_n$-Lipschitz extension of Definition~\ref{def:ext}; the right-hand side converges to zero in probability since $r_n\epsilon_n\rightarrow\infty$ and $(Z_n)$ is tight; this is where the lower bound on the numerical differencing step is used. Combining with Assumption~\ref{ass:unif} applied with $t_n=\epsilon_n$, we obtain, for every $c>0$, $\sup_{\|h\|\leq c}|\hat L^\prime_{\beta_n}(\theta,h)-L^\prime_n(h)|\rightarrow0$ in probability. Splitting on $\{\|\mathbb Z_n^\ast\|\leq c\}$ exactly as in Step~1, and using $\rho_n^{\mathbb Z}\rightarrow0$ together with the Lipschitz continuity of $L^\prime_n$ and the uniform tightness in probability of $(\mathbb Z_n^\ast)$, we conclude that
\begin{eqnarray}
\label{eq:rho-star}
\rho_n^\ast & := & \sup_{f\in\mbox{\tiny BL}_1(\mathbb R)}\left| \mathbb E\left[ f(\hat L^\prime_{\beta_n}(\theta,\mathbb Z_n^\ast)) \,\vert\, (Y_i,X_i)_{i=1}^n \right]
-\mathbb E\left[ f(L^\prime_n(\mathbb Z_0)) \right] \right| \; \rightarrow \; 0
\end{eqnarray}
in probability. 

\medskip\noindent\emph{Step 3: } Let $F_n$ denote the distribution function of $L^\prime_n(\mathbb Z_0)$, and fix $\delta>0$ small enough that the second display of Assumption~\ref{ass:cv}(b) holds, with $\gamma(\delta):=\liminf_n\left[(1-\alpha)-F_n(c_n-\delta)\right]>0$.

We first show that $\mathbb P(\hat c_n<c_n-\delta)\rightarrow0$. On the event $\{\hat c_n<c_n-\delta\}$, the definition of the conditional quantile gives $\mathbb P\left(\hat L^\prime_{\beta_n}(\theta,\mathbb Z_n^\ast)\leq c_n-\delta\,\vert\,\mbox{data}\right)\geq1-\alpha$, while the second inequality in~(\ref{eq:blsmooth}), applied conditionally on the data with $A=\hat L^\prime_{\beta_n}(\theta,\mathbb Z_n^\ast)$, $B=L^\prime_n(\mathbb Z_0)$ and step $\delta/2$, gives
\begin{eqnarray*}
\mathbb P\left(\hat L^\prime_{\beta_n}(\theta,\mathbb Z_n^\ast)\leq c_n-\delta\,\vert\,\mbox{data}\right)
& \leq & F_n(c_n-\delta/2) + (1+2/\delta)\rho_n^\ast \\
& \leq & 1-\alpha-\gamma(\delta/2)/2 + (1+2/\delta)\rho_n^\ast
\end{eqnarray*}
for all $n$ large enough. The event $\{\hat c_n<c_n-\delta\}$ therefore implies $\{(1+2/\delta)\rho_n^\ast\geq\gamma(\delta/2)/2\}$, whose probability vanishes by~(\ref{eq:rho-star}).

Next, using the first inequality in~(\ref{eq:blsmooth}) with $A=T_n$, $B=L^\prime_n(\mathbb Z_0)$ and step $\delta$,
\begin{eqnarray*}
\mathbb P\left( T_n>\hat c_n\right)
& \leq & \mathbb P\left( T_n>c_n-\delta\right) + \mathbb P\left( \hat c_n<c_n-\delta\right) \\
& \leq & 1-F_n(c_n-2\delta) + (1+1/\delta)\rho_n + \mathbb P\left( \hat c_n<c_n-\delta\right) \\
& \leq & \alpha + \left[ F_n(c_n)-F_n(c_n-2\delta)\right] + (1+1/\delta)\rho_n + \mathbb P\left( \hat c_n<c_n-\delta\right),
\end{eqnarray*}
where the last inequality uses $F_n(c_n)\geq1-\alpha$, which holds by definition of the quantile. Taking $\limsup_n$, the last two terms vanish by~(\ref{eq:rho-n}) and the previous paragraph, so that
\begin{eqnarray*}
\limsup_n\;\mathbb P\left( T_n>\hat c_n\right) & \leq & \alpha + \limsup_n\left[ F_n(c_n)-F_n(c_n-2\delta)\right],
\end{eqnarray*}
and letting $\delta\downarrow0$ and invoking the first display of Assumption~\ref{ass:cv}(b) gives $\limsup_n\mathbb P(T_n>\hat c_n)\leq\alpha$. In either case, $\limsup_n\mathbb P(E_n\,\vert\,(X_i)_{i=1}^\infty)\leq\alpha$ along the subsequence, almost surely, and Lemma~\ref{lem:transfer}(i) completes the proof.
\end{proof}

\section{Details of the Monte Carlo implementation}
\label{sec:simulation-details}

\subsection{Auxiliary scores and feasible covariance}

For either auxiliary model, let $w^B_{ij}$ and $w^S_i$ denote the regressor vectors in the big-firm and small-firm equations of Section~\ref{sec:simulation}, and write $\beta=(\beta_B^\top,\beta_S^\top)^\top$. The market-level estimating equations are
\begin{equation}
\label{eq:simulation-score}
s(Y_i\mid X_i;\beta)=
\begin{pmatrix}
\displaystyle \frac14\sum_{j\in G_B}w^B_{ij}
\bigl(Y_{ij}-w_{ij}^{B\top}\beta_B\bigr)\\[6pt]
\displaystyle w^S_i\bigl(N_{S,i}-w_i^{S\top}\beta_S\bigr)
\end{pmatrix}
=a_i-R_i\beta,
\end{equation}
where $R_i$ and $a_i$ are
\[
R_i=\begin{pmatrix}
\displaystyle \frac14\sum_{j\in G_B}w^B_{ij}w_{ij}^{B\top} & 0\\[6pt]
0 & w^S_iw_i^{S\top}
\end{pmatrix},
\qquad
a_i=\begin{pmatrix}
\displaystyle \frac14\sum_{j\in G_B}w^B_{ij}Y_{ij}\\[6pt]
\displaystyle w^S_iN_{S,i}
\end{pmatrix}.
\]
Thus, with $\bar R_n=n^{-1}\sum_iR_i$ and $\bar a_n=n^{-1}\sum_i a_i$, the OLS estimate satisfies $\bar R_n\hat\beta_n=\bar a_n$.

Since $\nabla_\beta s(Y_i\mid X_i;\beta)=-R_i$ does not depend on $Y_i$, Assumption~\ref{ass:uniq} holds automatically, including for the weighted estimating equations.

The score components differ in scale, and their scales change with the number of potential entrants. We therefore weight the estimating equations as in Appendix~\ref{sec:weights}, using a full covariance estimate constructed from five-fold cross-fitting as in Appendix~\ref{sec:cross-fit}. Let $f(i)$ denote market $i$'s fold, and define $m^B_j(X_i):=\mathbb E[Y_{ij}\mid X_i]$ and $m^S(X_i):=\mathbb E[N_{S,i}\mid X_i]$. The residual scores and covariance estimate are
\begin{equation}
\label{eq:simulation-covariance}
\hat\xi_i=
\begin{pmatrix}
\displaystyle \frac14\sum_{j\in G_B}w^B_{ij}
\bigl(Y_{ij}-\hat m^B_{j,-f(i)}(X_i)\bigr)\\[6pt]
\displaystyle w^S_i\bigl(N_{S,i}-\hat m^S_{-f(i)}(X_i)\bigr)
\end{pmatrix},
\qquad
\hat B_{c,n}=\frac1n\sum_{i=1}^n\hat\xi_i\hat\xi_i^\top.
\end{equation}
The conditional-mean estimates $\hat m^B_{j,-f(i)}$ and $\hat m^S_{-f(i)}$ use observations outside fold $f(i)$, and are obtained from an order-three Hermite polynomial sieve. Predictions in $\hat m^B_{j,-f(i)}$ and $\hat m^S_{-f(i)}$ are clipped to $[0,1]$ and $[0,J-4]$, respectively. 

For each sample and auxiliary model, we construct a whitening matrix $\hat W_n$ satisfying $\hat W_n\hat B_{c,n}\hat W_n^\top=I$, using a Cholesky decomposition of $\hat{B}_{c,n}$. 

\subsection{Equilibrium computation and numerical integration}

For a given candidate $\theta$, we compute the weighted binding function $L_n(\theta,\beta;\hat W_n)$ using the dual representation~\eqref{eq:dual} applied to $\hat W_ns$. This involves three steps: (1) solving the inner minimization
\begin{equation}
\label{eq:inner_problem_simu}
v_i(u,\lambda,\beta):=
\min_{y\in\Gamma(u\vert X_i;\theta_1)}
\lambda^\top\hat W_n s(y\vert X_i;\beta),
\end{equation}
(2) calculating its conditional expectation
\[
\mathbb E_{Q_{U\vert X_i;\theta_2}}
\bigl[v_i(U,\lambda,\beta)\mid X_i\bigr],
\]
and (3) solving the outer problem
\[
L_n(\theta,\beta;\hat W_n)
=\sup_{\lambda\in\mathcal S}
\frac1n\sum_{i=1}^n
\mathbb E_{Q_{U\vert X_i;\theta_2}}
\bigl[v_i(U,\lambda,\beta)\mid X_i\bigr].
\]
The implementation proceeds as follows.
\begin{enumerate}
\item \emph{Inner minimization.} Write $v_{ij}=\alpha X_{ij}+U_{ij}$. For each possible big-firm entry count $b\in\{0,\ldots,4\}$, sorting the small firms' payoff indices and checking their entry inequalities determines, almost surely under the continuous shock distribution, the unique compatible small-firm count, obtained by binary search. We then check the best-response inequalities for the $2^4=16$ possible big-firm action profiles. Because the auxiliary scores depend on small-firm entry only through its total count, small-firm identities need not be enumerated. 

\item \emph{Numerical integration.} We integrate one big-firm Gaussian shock analytically, and use randomized quasi-Monte Carlo integration for the remaining $J-1$ shocks. We use Golden-sequence points transformed by the standard-normal quantile to generate 100 quasi-random draws needed for the quasi-Monte Carlo integration.

\item \emph{Outer optimization.} Note that the objective function in the outer problem is concave in $\lambda$. Thus, one can solve it with a suitable convex optimization method, using a linear optimization oracle for the attainable-moment set. Here, we solve it through its equivalent minimum-distance representation.  Let $\mathcal A_{n,K}(\theta)$ be the simulated convex set of attainable raw outcome moments and $\mathcal C_{n,K}(\theta)=\hat W_n\mathcal A_{n,K}(\theta)$. Then, we have 
    \begin{equation*}
        L_n(\theta,\beta;\hat W_n) \approx \min_{\mu\in\mathcal C_{n,K}(\theta)}\bigl\|\mu-\hat W_n\bar R_n\beta\bigr\| = \operatorname{dist}(\hat W_n\bar R_n\beta, \mathcal{C}_{n,K}(\theta))
    \end{equation*}
    where the approximation reflects the possible error of numerical integration in the previous step. We then solve this minimum-distance problem using the fully-corrective Frank-Wolfe algorithm. This reduction to distance from a set of raw moments independent of $\beta$ need not hold for a general nonlinear auxiliary model. In such cases, the concave dual program~\eqref{eq:dual} can instead be solved using standard convex optimization methods.
\end{enumerate}

For the numerical delta step, write $\hat t_n=\hat W_n\bar a_n=\hat W_n\bar R_n\hat\beta_n$. For each sample and auxiliary model, we draw 500 standard Gaussian directions $z_b\in\mathbb R^{d_\beta}$, where $d_\beta=4$ or $7$, depending on which auxiliary model is used. With $\epsilon_n=2n^{-1/4}$, we compute
\[
D_{n,b}(\theta)=
\frac{
\operatorname{dist}\bigl(\hat t_n+\epsilon_nz_b,\mathcal C_{n,K}(\theta)\bigr)
-\operatorname{dist}\bigl(\hat t_n,\mathcal C_{n,K}(\theta)\bigr)
}{\epsilon_n}.
\]
This is numerically equivalent to perturbing the coefficient vector to $\hat\beta_n+\epsilon_n\mathbb Z^\ast_{n,b}$ with $\mathbb Z^\ast_{n,b}:=\bar R_n^{-1}\hat W_n^{-1}z_b$, whose law is $N(0,\hat\Omega_c)$, since $\nabla_\beta s=-R_i$ implies $\hat H_n=\bar R_n^{-1}$ and $\hat W_n^{-1}\hat W_n^{-\top}=\hat B_{c,n}$ in Appendix~\ref{sec:cross-fit}. The test statistic is $\sqrt{n}L_n(\theta,\hat\beta_n;\hat W_n)$, and its critical value is the $0.95$ quantile of the $D_{n,b}(\theta)$. 

\subsection{Table \ref{tab:simulation-projections} design}
The projection experiment in Table \ref{tab:simulation-projections} uses $100$ repeated experiments and a regular grid with $51$ points on each axis:
\[
\alpha\in[0.8,2.05],\qquad
\gamma_B\in[0.7,1.7],\qquad
\gamma_S\in[0,50/3].
\]
The spacings are $0.025$, $0.02$, and $1/3$, respectively, so that the true parameter $(1,1,4)$ is a regular grid point. These search spaces for $\alpha$, $\gamma_B$ and $\gamma_S$ strictly contain the confidence sets found in all repeated experiments.

\end{appendix}

\bibliographystyle{abbrvnat}
\bibliography{IndirectInference}

\end{document}